\documentclass[a4paper,12pt]{article}
\usepackage{jcappub} 
\usepackage{cite}
\usepackage{graphicx}
\usepackage{enumerate}
\usepackage{hyperref}

\usepackage{cancel}
\usepackage{color}
\usepackage{graphicx}
\usepackage{amsmath}
\usepackage{amsfonts}
\usepackage{amssymb}
\usepackage{rotating}
\usepackage{booktabs}
\usepackage{xcolor}
\usepackage{soul}

\def\blue#1{\textcolor{blue}{#1}}
\def\green#1{\textcolor{green}{#1}}

\def\comment#1{}

\usepackage{graphicx}

\title{\boldmath 
Gravo-thermal catastrophe in gravitational collapse and energy progenitor of Gamma-Ray Bursts
}

\author{She-Sheng Xue}

\affiliation{ICRANet Piazzale della Repubblica, 10 -65122, Pescara, Italy, \\ Physics Department, University of Rome La Sapienza, \\ P.le Aldo Moro 5, I–00185 Rome, Italy
}

\emailAdd{xue@icra.it and shesheng.xue@gmail.com} 

\abstract{We study the homologous collapse of stellar nuclear core, the virial theorem for hadron collisional relaxations, and photon productions from hadron collisions. We thus show the gravo-thermal dynamical process that transforms gravitational energy to photon energy. The process is energetically and entropically favourable. The total baryon number conservation, Euler equation for energy-momentum conservation and Poisson's equation for gravitational potential are adopted to describe homologous core collapses. The virial theorem determines the hadron collision energy gain from gravitational potential.  The hadronic photon production rate determines the photon energy density.  
The time scales of macroscopic and microscopic processes are studied to verify approximations. As a result, we show the formation of opaque photon-pair spheres, whose total energy, size, temperature and number density, accounting for the main energetic features of Gamma-Ray Burst progenitors. We obtain the intrinsic correlations of these quantities. 
They depend only on the averaged thermal index of the stellar core. We discuss the possibility to confront them with observational data.
}

\begin{document}

\maketitle
\flushbottom

\newpage

\section{Introduction}\label{sec:0}

Gamma-Ray Bursts (GRBs) are the most energetic and complex events in the Universe. One of the peculiar features is that the progenitors of these events release hard photons of tremendous energy $10^{49}\sim 10^{54}$ ergs in a few seconds. 
Since the last decades, great progress has
been made for understanding the GRB natures \cite{2004RvMP...76.1143P,2006RPPh...69.2259M,2014ARA&A..52...43B,2015JHEAp...7...73D,2015PhR...561....1K,zhang_2018,Ruffini2019,2019Univ....5..110R,2021MNRAS.504.5301R}. 
However it still remains an open question what is the dominate dynamics of GRB progenitors, and many efforts have been made to find answers  \cite{1993ApJ...405..273W,1998ApJ...502L...9F,1998Natur.395..663K,1999ApJ...524..262M,Preparata1998,
Preparata2003,Ruffini2008,Ruffini2010,Han2012,Ruffini2013,
2002AJ....123.1111B,2007ARA&A..45..177C,2006ApJ...637..914W,2006Natur.444.1047F,2016ApJ...821L..18P,2018ApJ...857..128J,2000A&A...359..855R,2001ApJ...556L..37M,2003nvm..conf..185D,2011PhRvL.106y9001W,2013ApJ...778...67N,2014ApJ...793L..36F,2016ApJ...832..136R,2018ApJ...854...43H,Li2018,Ruffini2019}. 
It is no doubt that massive stellar gravitational potentials are great energy reservoirs, stellar collapse and coalescence must gain gravitational energies. How huge and invisible potential energies are converted to visible hard photon energies in just a few seconds, accounting for GRBs. Slowly hydrodynamical and kinematic processes are probably unlikely to make such rapid energy conversions. The most probable candidatures are electromagnetic and/or strong interacting processes. On the latter candidature, we mostly focus on this article.


In a self-gravitating system, Lynden-Bell and Wood \cite{LyndenBell1967, LyndenBell1968} pioneered the study of gravo-thermal catastrophe by using violent particle collisionless relaxation and virial theorem. The relevant dynamics and time scales in galactic and stellar systems have been discussed in the references, see for example \cite{Padmanabhan1989, Chavanis2002c, Chavanis2003, Sormani2013,Roupas2015,Roupas2017, Takacs2018, Dutta2019}. We are inspired by the Lynden-Bell and Wood study, which shows the relaxation process that particles gain their mean kinetic heat energy from gravitational potential. On the other hand, we learn the hadron and quark-gluon photon productions by heavy-ion collisions. In the last decades, such photon productions have been intensively studied theoretically and experimentally in nuclear and particle physics \cite{Turbide2004, Kapusta1991, Kapusta1992,
Arnold2001,Gale2015,Hidaka2015}. This microscopic photon production process may have important applications in the arena of astrophysics and cosmology. 

Suppose that hadrons inside compact stellar core gain gravitation energy by collisional relaxation, and hadron collisions produce photons. We study the gravo-thermal dynamics: how the gravitational core-collapse process converts the gravitational energy to photon energy and create a photon-pair sphere. To qualitatively show how the dynamics work, we adopt homologous core collapses \cite{Goldreich1980} and discuss different time and length scales of macroscopic and microscopic processes. Using both analytical and numerical approaches for calculations, we show the formation of opaque and energetic photon-pair spheres, possibly explaining GRB progenitors.
 
We organise this article as follow. In Secs.~\ref{virial} and \ref{hadron}, we present the discussions of gravo-thermal catastrophe, virial theorem, and photon productions from hadron collisions. We discuss the gravo-thermal dynamics by using homologously collapsing core in Secs.~\ref{Homo} and \ref{sec:1}.
In Secs.~\ref{psphere} and \ref{scaling}, we analyse photon-pair sphere properties and intrinsic correlations in connection with GRB progenitor features. Finally, we show that photon-pair sphere formation is energetically and entropically favourable in 
Sec.~\ref{form}, and give conclusion and remarks 
in Sec.~\ref{end}. The stellar core density is characterised by the nuclear saturation density $n_0\approx 0.16 \,{\rm fm}^{-3}=1.6\times 10^{38}/{\rm cm}^3$ or $\rho_0=mn_0\approx 2.4\times 10^{35} {\rm ergs}/{\rm cm}^3$, and $m\approx 1$ GeV is the typical baryon (neutron) mass. The light speed $c=1$, Planck constant $\hbar =1$ and Boltzmann constant $k=1$ are used, unless otherwise specified.  

\section{Gravo-thermal catastrophe and virial theorem}\label{virial}

A stellar core is composed of nuclear matter, whose total mass $M$ and baryon number $N=M/m$ of baryons of mass $m$ \footnote{By the term {\it baryons}, we indicate nucleons, nuclei, hadrons, quark-gluon plasma that carry baryon numbers.}. 
This is a self-gravitating system. Its negative gravitational energy $U$ 
and potential $\phi$,
\begin{eqnarray}
U=\frac{1}{2}\int dx^3 \rho(x)\phi <0,\quad
\phi(x)=- G\int d^3x'\frac{\rho(x)}{|x-x'|},
\label{gravPU}
\end{eqnarray}
where $G$ is the Newton constant. In the time-varying gravitational potential, 
via collisionless relaxation, baryons gain their
mean kinetic energy from the gravitational energy. 
The baryon mean kinetic energy contributes to 
the core internal energy $F$, that in turn balances 
the self-gravitating potential energy. 
Such gravo-thermal catastrophe phenomenon of stellar cores is studied by considering violent relaxation \cite{LyndenBell1967} and 
equipartition theorem \cite{LyndenBell1968}. In these studies, they obtained the collisionless relaxation timescale $(3/4)(2\pi G \bar \rho)$ of baryons in time-varying gravitational potential, in terms of the core mean density $\bar \rho$. 
The stellar core is assumed to be an isothermal core of temperature $T\ll m$ and internal ``heat" energy $F=\frac{3}{2}TN\ll mN$. Moreover, they considered the isothermal core as an equilibrium or equipartition system.
Therefore, the virial theorem of Clausius can be applied
\begin{eqnarray}
2F+ U =3 PV,
\label{gravP} 
\end{eqnarray}
where $V$ is the core volume and $P$ is the external pressure, acting on the core surface.

These studies show that the baryon collisionless relaxation process in 
gravitational potential and external pressure leads to two consequences. 
(i) The gravitational energy is converted to the internal heat energy of baryon gas. 
(ii) The balance between gravitational dynamics and thermodynamics is established, 
and the virial theorem (\ref{gravP}) is applicable. 
The second point is correct, provided the time scales of relaxation processes 
is much smaller than that of macroscopic gravitational and hydrodynamical processes.

For the reasons given in the next paragraph, we generalise these discussions from an entire 
stellar core to a small fluid element of volume $dV$ inside the stellar core. 
The virial theorem (\ref{gravP}) can be given by
\begin{eqnarray}
2dF+ dU &=&3 pdV,\quad dF=(3/2)T (\rho/m)dV,\quad dU=(1/2)\rho \phi dV,
\label{dgravP} 
\end{eqnarray}
and $dF\ll \rho dV$, i.e., $T\ll m$. 
Associating to each fluid element 
$\rho dV$, the baryon temperature $T$ characterises the mean kinetic 
energy of baryon motions and collisions.  
We approximately describe the internal pressure $p$  by the equation of state (EoS)
\begin{eqnarray}
p = \kappa \rho ^\gamma,
\label{eos} 
\end{eqnarray}
with two parameters: mean thermal index $\gamma$ and 
coefficient $\kappa$. The latter relates to the maximal value of 
core centre density $\rho_c$. Here we have to mention that the EoS (\ref{eos}) 
is adopted for this simple model and preliminary study. Because of large core density and possible phase transition, more 
complex types of EoS are expected, for example, $\gamma$ and 
$\kappa$ are not constants in space and time.  

We will consider compact stellar cores of nuclear density so that equipartition or equilibrium should be achieved by the relaxation processes of baryon-baryon collisions. 
Unlike the virial theorem (\ref{gravP}) based on collisionless relaxation, 
the virial theorem (\ref{dgravP}) mainly bases on collisional relaxation. 
The collisional relaxation timescale can be estimated as
\begin{eqnarray}
\tau_{\rm relax}=(\sigma_n v n)^{-1}\approx 5.21\times 10^{-3} (n_0/n) (T/m)^{1/2}T^{-1} < T^{-1},
\label{relax} 
\end{eqnarray}
where $n=\rho/m$ ($v$) is the baryon number 
density (mean velocity), $\sigma_n\approx 1/m_\pi^2$ is the typical cross-section of baryon-baryon collisions, 
and averaged collision energy (baryon temperature) $T=(1/2)mv^2\ll mc^2$.
The baryon collisional relaxation time scale $\tau_{\rm relax}$ (\ref{relax}) 
is smaller than the time scales ($T^{-1}$) of baryon kinetic motion and collisionless relaxation \footnote{In 
Eq.~(\ref{relax}), for $T\sim 10^2$ MeV and $n\sim n_0$, 
the baryon kinetic motion time scale $T^{-1}\sim 10^{-23}$ seconds, and the baryon collisional relaxation time scale (\ref{relax}) $\tau_{\rm relax}\sim 10^{-26}$ seconds.}. 
Therefore, a {\it local} equilibrium or equipartition state
can be established. Thus, the {\it local} virial theorem (\ref{dgravP}) is applicable. 
As a consequence, the baryon temperature $T$ is determined by 
\begin{eqnarray}
3T/m &\approx & 3v_s^2 (1/\gamma) - (1/2) \phi,
\label{dgravT}\\
v_s^2 &=&\partial p/\partial \rho =\gamma p/\rho\nonumber
\end{eqnarray}
where $v_s$ is the sound velocity. 
For an isothermal core, integrating 
Eq.~(\ref{dgravP}) over the core volume $\int dV$, one obtains the virial theorem (\ref{gravP}), 
using $\int pdV =PV$.

\section{Photon production from hadron collisions}\label{hadron}

Through baryon-baryon collisions, 
baryons gain their mean kinetic (heat) energy from the gravitational potential of self-gravitating cores. 
The mean baryon collision energy is characterised by the baryon temperature $T$ (\ref{dgravT}). 
Besides, the baryon-baryon collisions produce photons and pairs of light charged leptons and quarks, 
As a result, baryons' mean kinetic heat energy is converted to the energy of photons, 
and pairs of other light charged 
leptons and quarks.  

From the studies of heavy-ion collisions, see for example references \cite{Turbide2004, Kapusta1991, Kapusta1992,
Arnold2001,Gale2015,Hidaka2015}, 
we learn the hadron productions of photons in the heavy-ion collision.
References \cite{Kapusta1991, Kapusta1992} show high-energy photons produced from quark-gluon plasma versus hot hadronic gas. The photon production rates of the quark-gluon plasma and the hadron gas are approximately the same in the energy range of a few hundred MeV. The transition from hadron matter to quark-gluon plasma (QGP) occurs at a temperature of about 200 MeV. In this energy range, quarks and gluons are deconfined in colourless particles, and 
become the pertinent degrees of freedom of the system. 
Given the temperature $T$ of hadron gas (or quark gluon plasma) collisions, the photon production rate
$R_\gamma =dN_\gamma/dtdV$ per volume is \cite{Turbide2004}
\begin{equation}
    \frac{dR_\gamma}{d^3{\vec q}}= q^{-1}_0 \frac{2}{3}\frac{\alpha\alpha_s}{2\pi^2}T^2 e^{-q_0/T}\ln\left( 1+ \frac{2.9}{4\pi\alpha_s}\frac{q_0}{T}\right),
    \label{gammaR}
\end{equation}
where QED and QCD fine structure constants $\alpha=1/137$ and $\alpha_s\approx 0.5$. 

Based on the hadronic photon production (\ref{gammaR}) in heavy-ion collisions, 
the production rates of the photon number and energy densities can be calculated as, 
\begin{equation}
\frac{dn_\gamma}{dt} =  \int d^3{\vec q} ~\frac{dR_\gamma}{d^3q},
\quad {\rm and} \quad \frac{d\rho_\gamma}{dt} 
= \int d^3{\vec q}~ q_0 ~\frac{dR_\gamma}{d^3q}.
\label{nerate}
\end{equation}
In addition to photons, we consider pairs of relativistic charged leptons or quarks are 
approximately massless, i.e., $q^2=q_0^2-{\vec q}^2=0$ at energy scale $T$ 
of a few hundred MeV. Their chemical potentials are zero, due to lepton- or baryon-number conservation. 
Integrating over the phase space $d^3{\vec q}=4\pi q^2 dq$ in 
Eq.~(\ref{gammaR}), 
we obtain, 
\begin{equation}
    R_\gamma \approx  \frac{4}{3}\frac{\alpha\alpha_s}{\pi}T^4\ln\left( 1+ \frac{2.9}{4\pi\alpha_s}\right).
    \label{gammaNR}
\end{equation}
Here we use the saddle point approximation $q_0\approx T$ to arrive at a simple formula for astrophysical applications.  
The exponential suppression factor $e^{-q_0/T}$ (\ref{gammaR}) shows the dominant photon production at 
energy $q_0\sim T$. On the other hand, the Heisenberg uncertainty relation 
in thermal particle productions yields $q_0 \tau\sim 1$. This implies that the thermal photon production time scale $\tau_{\rm prod}\sim q^{-1}_0\sim T^{-1}$. Therefore, the photon number and energy densities (\ref{nerate}) can be approximately estimated as,
\begin{eqnarray}
n_\gamma &\approx & \frac{4}{3}\frac{\alpha\alpha_s}{\pi}T^3\ln\left( 1+ \frac{2.9}{4\pi\alpha_s}\right),
\label{gammaN}\\
\rho_\gamma &\approx & \frac{8}{3}\frac{\alpha\alpha_s}{\pi}T^4\ln\left( 1+ \frac{2.9}{4\pi\alpha_s}\right).
\label{rhog}
\end{eqnarray}
We thus consistently define the production time scale 
\begin{eqnarray}
\tau_{\rm prod}\equiv n_\gamma/R_\gamma \approx T^{-1} > \tau_{\rm relax}.
\label{prodt}
\end{eqnarray}
These are qualitative properties of photons produced by heavy-ion collisions at baryon temperature $T$. 

\comment{
Similarly, from Eq.~(\ref{gammaR}), the production rate of the photon energy per volume $\rho_\gamma$ is obtained 
\begin{equation}
\frac{d\rho_\gamma}{dt} =  \int d^3{\vec q} q_0\frac{dR_\gamma}{d^3q} \approx  \frac{8}{3}\frac{\alpha\alpha_s}{\pi}T^4\ln\left( 1+ \frac{2.9}{4\pi\alpha_s}\right),
    \label{gammaE}
\end{equation}
using $\int_0^\infty 4\pi q^2 dq e^{-q/T}= 8\pi T^3$. }

The photon productions by baryon-baryon collisions in an isothermal baryon core of temperature $T$ (\ref{dgravT}) 
are inevitable. The necessary condition is that photons are massless, and no energy gap needs to overcome in their thermal production. Other relativistic charged lepton or quark pairs are approximately massless for temperature being much larger than their masses ($T\gg m_{\ell,q}$). 
The sufficient condition on the other hand is that the photon production time scale $\tau_{\rm prod}$ (\ref{prodt}) should be larger than the relaxation time scale 
$\tau_{\rm relax}$ (\ref{relax}). This is indeed the case 
$\tau_{\rm prod} > \tau_{\rm relax}$. 
The gravo-thermal catastrophe produces not only massive baryons' heat energy 
$(3/2)T(\rho/m) dV$ (\ref{dgravP}), but also relativistic particles' energy $\rho_\gamma dV$ (\ref{rhog}).

For the baryon temperature $T\sim {\mathcal O}(10^2)$ MeV, 
the photon density $n_\gamma$ (\ref{rhog}) is so large that relativistic particles are opaqued and thermalized by collisions among themselves. The photon thermilization time scale, which is the mean-free path divided by the speed of light $c$, can be estimated by
\begin{eqnarray}
\tau_{\rm therm}=(\sigma_\gamma c n_\gamma)^{-1}\approx \frac{3.23\times 10^{6}}{[\ln (2T_\gamma/m_e)]} \left(\frac{m_eT_\gamma}{T^2}\right) \frac{1}{T} \gg \tau_{\rm prod}.  
\label{Tgamma}
\end{eqnarray} 
Here we use the Thomson cross section $\sigma_\gamma \approx \sigma_T(3/8)(m_e/T_\gamma)\ln (2T_\gamma/m_e)$ and  
$\sigma_T=(8\pi/3)\alpha^2/m^2$. The photon temperature 
$T_\gamma$ represents the mean energy of photon collisions.
The mean temperature $T_\gamma$ of thermalized photons can be estimated by equating 
photon thermal energy density to Eq.~(\ref{rhog})
\begin{eqnarray}
\frac{\pi^2}{15}T_\gamma^4\approx \rho_\gamma,\quad
\Rightarrow \quad T_\gamma \approx 0.21 T.  
\label{Tphoton}
\end{eqnarray} 
It shows that the photon temperature $T_\gamma$ is the same order of the 
baryon temperature $T$. 
Inequality $\tau_{\rm therm} \gg \tau_{\rm prod}$ (\ref{Tgamma}) implies that many photons are produced by baryon collisions before they are  thermalised through their electromagnetic interactions. We further note that
the viscous hydrodynamical evolutions of photons start at thermalisation  
time $\tau_{\rm therm}$ 
\cite{Turbide2004,Vujanovic2014}.  

Moreover photons produced from baryon-baryon collisions electromagnetically are coupled to pairs of charged leptons (quarks) and anti-leptons (-quarks). Among these charged particle pairs, we particularly consider pairs of electrons and positrons, since their mass (MeV) is smaller than the photon characteristic energy $\sim 
T > $ MeV (\ref{dgravT}). Therefore the forth and back processes 
$\gamma+\gamma\leftrightarrow e^++e^-$ of electron and positron pair production and annihilation are kinematically possible. The cross section $\sigma_{\gamma\gamma\leftrightarrow e^+e^-}$ 
is of the order of the Thomson cross section $\sigma_\gamma$ in Eq.~(\ref{Tphoton}), and the corresponding mean-free length 
$\sim (\sigma_{\gamma\gamma\leftrightarrow e^+e^-}n_\gamma)^{-1}$ 
is very small. This indicates that via processes $\gamma+\gamma\leftrightarrow e^++e^-$, photons and electron-positron 
pairs participate thermalisation or thermal equipartition in particle energy and number. 
In addition, electron-positron pairs can be produced by photons interacting with 
	viral photons $\gamma^*$ of classical external magnetic fields \cite{Daugherty1975,1983ApJ...273..761D}. 
	Since the electron and positron pairs are relativistic at the energy scale under consideration, 
the degree freedoms of photon polarisation and electron-positron pair are the same, 
their thermal equipartition results in the particle energy densities $\rho_{e^+e^-}\approx \rho_\gamma$ and number densities $n_{e^+e^-}\approx n_\gamma$, conserving the total lepton number.  
Thus, a half of photons produced from baryon-baryon collisions converts to electron-positron pairs. Equation (\ref{Tphoton}) 
becomes $(\pi^2/15)(T_\gamma^4+T_{e^+e^-}^4)\approx \rho_\gamma$, where $T_\gamma\approx T_{e^+e^-}$. 
The same discussions can be applied to the inclusion of the lightest $u$ and $d$ quarks if the mean photon energy can excite them to participate in the thermalisation. As a result, these processes lead to an electrically neutral, deeply opaque and thermalised plasma (fluid) of photons and electron-positron pairs. 

To end this section, it is worthwhile to have a brief discussion on other possibilities of photon productions from baryon-baryon collisions. We point out that the hadronic photon production (\ref{gammaR}) or (\ref{nerate}) is proportional to 
$\alpha\alpha_s$. The strong QCD coupling constant $\alpha_s$ comes from the kinematic motion of charged quark-gluon matter (fluid) inside hadrons, which is predominately governed by the shorted ranged QCD dynamics. Whereas, the QED coupling constant $\alpha$ 
come from the charged quark-gluon matter coupling to photons, and producing photons via its QCD govern kinematic motion. 
It is conceivable that the kinematic motions of charged baryonic matter are partially governed by the long-ranged QED dynamics, and the photon production proceeds via QED processes like thermal bremsstrahlung, double Compton scattering, and Comptonized cyclotron/synchrotron in the presence of fluctuating electromagnetic fields in microscopic length scales and/or classical magnetic fields in macroscopic length scales. In this case,
the photon production rates should be proportional to at least $\alpha\alpha$, which is smaller than $\alpha\alpha_s$. The reason for $\alpha\alpha$ is that both charged baryonic particle motions and radiations are attributed to the QED dynamics. 
These preliminary and qualitative discussions imply that 
the QCD hadronic photon-production rate (\ref{gammaR}) or (\ref{nerate}) should be the leading order contribution, while the contributions from the QED photon-production processes should be the next leading order. Detailed and quantitative calculations should be possibly performed by simulations in heavy-ion collision physics. 
Nevertheless, the contributions from QED photon-production processes further increase the photon-production efficiency in baryon-baryon collisions. As a consequence, the total number density of produced photons should be larger than the one (\ref{gammaN}) from the QCD process. These QED processes are in favour of increasing the energy and number densities of thermalised plasma (fluid) of photons and electron-positron pairs, which we discussed above.

\section{Adiabatic process of gravitational collapse}\label{Homo} 

We now turn to study the gravo-thermal dynamics. How gravitational 
collapses convert the gravitational potential energy to the photon
sphere energy via baryon collisional relaxation in gravitational 
potential, and baryon-baryon collisions for photon productions. 
The problems are that gravitational collapse is a dynamical avalanche process instead of a stationary process for which we discuss the 
virial theorem for particle collisionless relaxation 
in Sec.~\ref{virial}. Besides, baryon collisional relaxation and photon production are dynamical back reaction processes, as aforementioned. In practice, dealing with such a complex system, 
it is rather difficult for both analytical 
and numerical approaches to analyse all dynamics without any approximation. 
To make appropriate approximations, 
we need to understand what is dominant physical process and variation 
in given relevant length and time scales.   

\subsection{Fundamental equations for gravitational collapse}

Following the approach to homologously gravitational collapsing \cite{Goldreich1980}, 
we adopt in Newtonian approximation the continuity equation for total baryon number 
conservation, Euler equation for energy-momentum conservation and Poisson's equation for 
gravitational potential 
$\phi$, 
\begin{eqnarray}
  \frac{\partial \rho}{\partial t} +{\bf \nabla}\cdot(\rho \bf {u})&=& 0, \label{con}\\
   \Big(\frac{\partial \rho}{\partial t}\Big)+ {\bf \nabla}\Big(\frac{1}{2}|{\bf u}|^2\Big) + (\nabla\times {\bf u})\times {\bf u} + \nabla h 
+ {\bf \nabla}\phi &=& 0, \label{eulerT}\\
  \nabla^2 \phi -4\pi G \rho &=& 0.\label{potenT}
\end{eqnarray}
Here, ${\bf u}$ is the baryon fluid velocity.  
Equation of State (\ref{eos}) yields the baryon heat function $h=H/\rho$ 
($H$ for Enthalpy)
\begin{eqnarray}
\nabla h = \nabla p/\rho;  \quad h = \int \nabla p/\rho = \frac{\kappa \gamma}{\gamma -1}\rho^{\gamma -1}.
\label{heat} 
\end{eqnarray}
The gradient of heat function gives a thermal force against gravity. 
If the baryon fluid flow is vorticity free, the velocity can be obtained 
from a stream function ${\bf u}=\nabla v$ up to a constant. 
The Euler equation (\ref{eulerT}) becomes
\begin{equation}
    \frac{\partial\upsilon}{\partial t}+\frac{1}{2}|{\bf \nabla} \upsilon|^2+h+\phi=0.
    \label{euler1T}
\end{equation}
In the following, we present the solution for arbitrary $\gamma$ value, generalising the $\gamma=4/3$ solution
\cite{Goldreich1980}.
Note that for $\gamma < 4/3$, self-gravitating cores become dynamically unstable, and undergo gravitationally homologous collapses, see for example Refs.~\cite{Riper1981, Cao2009c, Lou2011}.

\subsection{Macroscopic and microscopic time scale hierarchy}

Self-gravitating core collapse processes have the time scale 
$\tau_{\rm grav}= |{\bf R}|/|\dot {\bf R}|$, and $|{\bf R}|$ is the core radius.  
The notation $\tau_{\rm grav}$ represents also the time scale 
$\sim |{\bf R}|/v_s$ of other hydrodynamical processes, where $v_s$ is the sound velocity.
As will be shown in next selection \ref{sec:1}, the $\tau_{\rm grav}$ is a macroscopic 
time scale. It is much larger than the microscopic process time scales of baryon 
collisional relaxation $\tau_{\rm relax}$ (\ref{relax}), photon production $\tau_{\rm prod}$ (\ref{prodt}) and thermalization $\tau_{\rm therm}$ (\ref{Tgamma}), 
\begin{eqnarray}
\tau_{\rm grav}\gg \tau_{\rm therm}>\tau_{\rm prod}> \tau_{\rm relax}.
\label{Thy}
\end{eqnarray}
This time scale hierarchy means that (i) these microscopic processes 
occur not only ``instantaneously'', but also ``locally'' 
($\ell_{\rm micro} \approx  c\tau_{\rm micro}\ll |{\bf R}|$), 
comparing with gravitational collapsing 
and hydrodynamical process; (ii) there is no causal correlation among the concurrence 
of these microscopic processes at different space-time points in macroscopic scales. 
In other words, the vast difference in microscopic and macroscopic time scales 
implies that gravitational collapse can be considered as a very slowly 
{\it adiabatic} process, comparing with these rapid and local microscopic processes. 
Therefore, we regard that these microscopic processes can be 
approximately analysed, as if self-gravitating cores were static. Namely,  
in a gravitational collapse process, we can approximately adopt the results 
of baryon temperature $T$ (\ref{dgravT}) and photon temperature $T_\gamma$ (\ref{Tphoton}), 
as well as photon production number and energy densities (\ref{gammaN},\ref{rhog}), obtained in previously sections.
It is what we call the {\it adiabatic} approximation. 

Moreover, the shorter time scale process proceeds several times and has been well established in the smooth varying period of longer time scale processes. 
But the inverse is not correct. 
The time scale hierarchy (\ref{Thy}) shows that the most rapidly established microscopic process is the baryon collisional relaxation (\ref{dgravT}). The photon production process (\ref{prodt}) is less rapid. The photon thermalisation process (\ref{Tgamma}) is the slowest one.  These microscopic processes are well established within the smooth varying periods of gravitational collapse and hydrodynamical processes.
This time sequence (\ref{Thy}) provides a necessary condition for the gravo-thermal dynamics. Namely, from gravitational potential, baryons gain heat energy via their collisional relaxations (\ref{dgravP}). Via hadronic photon productions, baryons convert their heat energy 
to photon energy (\ref{gammaNR}), which are then 
thermalised (\ref{Tgamma}). 
  
{\it A priori}, we give the reasons why such an {\it adiabatic} approximation can be adopted to study microscopic dynamics in gravitational collapsing and hydrodynamic processes.
However, its self-consistency has to be verified {\it a posterior}.
It follows the same reasons, discussed in the pioneer works \cite{LyndenBell1975a, Shapiro1977a}. 
There it is stated that the gravitational evolution time scale due to 
stellar evaporation and consumption is longer than the core relaxation time scale, the core will maintain an approximate isothermal core profile (\ref{gravP}) 
and will evolve homologously. 

\comment{
This vast difference in micro and macroscopic time scales gives us a chance to track the problem analytically, though it is a problem of a numerical algorithm. 
We can think of the following scenario. Gravitational collapse establishes an energy equipartition state, i.e., virial theorem guarantees. 
The kinetic energy $F$ gains from the gravitation energy, and the virial theorem (\ref{gravP}) $2F=-U$, and gravothermal catastrophe $F=3/2  N (kT)$. The relaxation time of the process building up virial theorem is much faster, shorter than the gravitational collapse time scale $\dot R/R$. Therefore, such the virial relation $2F=-U$ and temperature $T$ built can apply. Besides, the baryon number is conserved. So we adopt the study of thermal Thermal photon and lepton pair production in heavy-ion collision.
1. $\gamma <4/3$ collapse, 2. no photons at the beginning and produced later by baryon collisions. 3. photon energy density is negligible compared with baryon mass density, so its role will not be considered in collapse, which will be discussed later. 4. photons are optic thick .}

\section{Homologously collapsing stellar core}\label{sec:1}

\subsection{Basic gravitational length scale and sound velocity}

We consider only the case of spherically symmetric stellar cores. 
The core radial coordinate is rescaled from (dimensioned) ${\bf R}$ to (dimensionless) ${\bf r}={\bf R}/a(t)$. 
The time dependent length scale function $a(t)$ is defined as
\begin{equation}
    a(t)=(\gamma p_{c}/\rho_{c})^{1/2}/(\gamma\pi G \rho_{c})^{1/2}=\rho^{\frac{\gamma}{2}-1}_{c}(\frac{\kappa}{\pi G})^{1/2},
		\label{a0}
\end{equation}
where the subscript or superscript $c$ indicates the values 
at core center ${\bf R}=0$ or ${\bf r}=0$. The baryon density is rewritten as
\begin{equation}
    \rho=\rho_{c}f^3,
		\quad 
		\rho_c=(\frac{\kappa}{\pi G})^{\frac{1}{2-\gamma}}
		a^{\frac{2}{\gamma-2}},
		\label{den1}
\end{equation}
where the core centre density $\rho_c=\rho_c(t)$ is a function of time, and homologous profile $f=f(r)$ is a function of dimensionless 
radius $r$ only. 
The scale function $a(t)$ decreases and centre density $\rho_{c}(t)$ increases
as gravitational collapse goes on. The gravitational potential $\phi$ 
is re-expressed 
\begin{equation}
    \phi =(\frac{\gamma p_{c}}{\rho_{c}})\psi =(v_s^c)^2\psi
		=\gamma \kappa \rho_c^{\gamma -1}\psi,
		\label{phi1}
\end{equation}
in terms of another homologous function $\psi$ and the sound velocity 
\begin{equation}
    v_s^2=\gamma p/\rho = (v_s^c)^2 f^{3(\gamma -1)}; \quad (v_s^c)^2\equiv\gamma p_c/\rho_c= \gamma\kappa \rho_c^{\gamma-1}.
\label{sound}
\end{equation}
The sound velocity at core center $(v_s^c)^2=\gamma p_c/\rho_c$ increases as the centre density $\rho_c$ increases and scale function $a(t)$ decreases. The maximal sound velocity $(v_s^c)^2=\gamma \kappa \rho_c^{\gamma-1}\leq 1/3$ \cite{Bedaque2015} leads to the minimal scale length $a_{\rm min}$ and maximal center density $\rho^{\rm max}_c$
\begin{equation}
a_{\rm min}= (3\gamma)^{\frac{2-\gamma}{2(\gamma-1)}}\Big(\frac{\kappa^{\frac{1}{\gamma-1}}}{\pi G}\Big)^{1/2},\quad \rho^{\rm max}_c= (3\gamma\kappa)^{-\frac{1}{\gamma-1}}.
\label{amini}
\end{equation}
The sound velocity and baryon density can be expressed as 
\begin{equation}
\rho_c=\rho_c^{\rm max} \Big(\frac{a}{a_{\rm min}}\Big)^{-\frac{2}{2-\gamma}};  \quad (v_s^c)^2= \frac{1}{3}\Big(\frac{a}{a_{\rm min}}\Big)^{-2\frac{\gamma-1}{2-\gamma}}=\frac{1}{3}\Big(\frac{\rho_c}{\rho_c^{\rm max}}\Big)^{\gamma-1}.
\label{soundc}
\end{equation}
These relates three functions $a(t)$, $(v_s^c)^2$ and $\rho_c$ at the core centre. We will use the centre density $\rho_c$ as 
the primal variable. The maximal core centre density $\rho_c^{\rm max}$ (\ref{amini}) 
is about $5\sim 8$ times the nuclear saturation energy density $\rho_0$. 
To illustrate numerical results in this article
we will adopt $\rho_c^{\rm max}=10\rho_0$. Thus, the coefficient $\kappa$ 
of EoS (\ref{eos}) is constrained. The only one parameter left is 
the averaged thermal index $\gamma$ in EoS (\ref{eos}). 

To present physical results and their relevance, we define the basic  gravitational length in homologous collapse:
\begin{eqnarray}
a_{\rm min} &=& 3.06\times 10^{5} (3\gamma)^{-1/2} (\frac{\rho_c^{\rm max}}{\rho_0})^{-1/2}  ({\rm cm}).
\label{amin}
\end{eqnarray}
The minimal scale length $a_{\rm min}$, ``sound horizon'', comes from Eq.~(\ref{amini}). The scale $a_{\rm min}$ gives the macroscopic length scale that we discussed in the time scale hierarchy (\ref{Thy}).

\subsection{Eigenvalue problem for homologous collapse configurations}

Setting ${\bf u}=\dot{a} {\bf r}$ or $v=(1/2)a\dot ar^2$ \cite{Goldreich1980}, we reduce the continuity equation (\ref{con}) to the trivial relation $\dot{f}=0$ which states that the homologous profile $f$ and density profile $f^3$ (\ref{den1}) do not evolve in time. 
Euler's equation (\ref{euler1T}) is separable in spatial ${\bf r}$ and temporal $t$ variables,
\begin{eqnarray}
\psi &=&\frac{1}{\gamma -1}\left(\frac{\lambda}{6}~ r^2-f^{3(\gamma -1)}\right), 
\label{psi}\\
a^{\frac{\gamma}{2-\gamma}}\ddot a &=& -\frac{\lambda}{6} \frac{2\gamma}{\gamma-1}\left(\frac{\kappa^{\frac{1}{\gamma-1}}}{\pi G}\right)^{\frac{\gamma-1}{2-\gamma}},
	\label{at}
\end{eqnarray}
and $\lambda$ is the eigenvalue. Via Eqs.~(\ref{phi1}) and (\ref{psi}), Poisson's equation (\ref{potenT}) yields the nonlinear eigenvalue equation for the homologous profile $f(r)$, 
\begin{equation}
    \frac{1}{r^2}\frac{d}{dr}\left(r^2 \frac{d f^{3(\gamma -1)}}{dr}\right)
		+\frac{4(\gamma-1)}{\gamma}f^3=\lambda,
		\label{e-profile}
\end{equation}
and the boundary conditions are $f'(0)=0$ and $f(0)=1$. 

\begin{figure}
\includegraphics[width=3.0in]{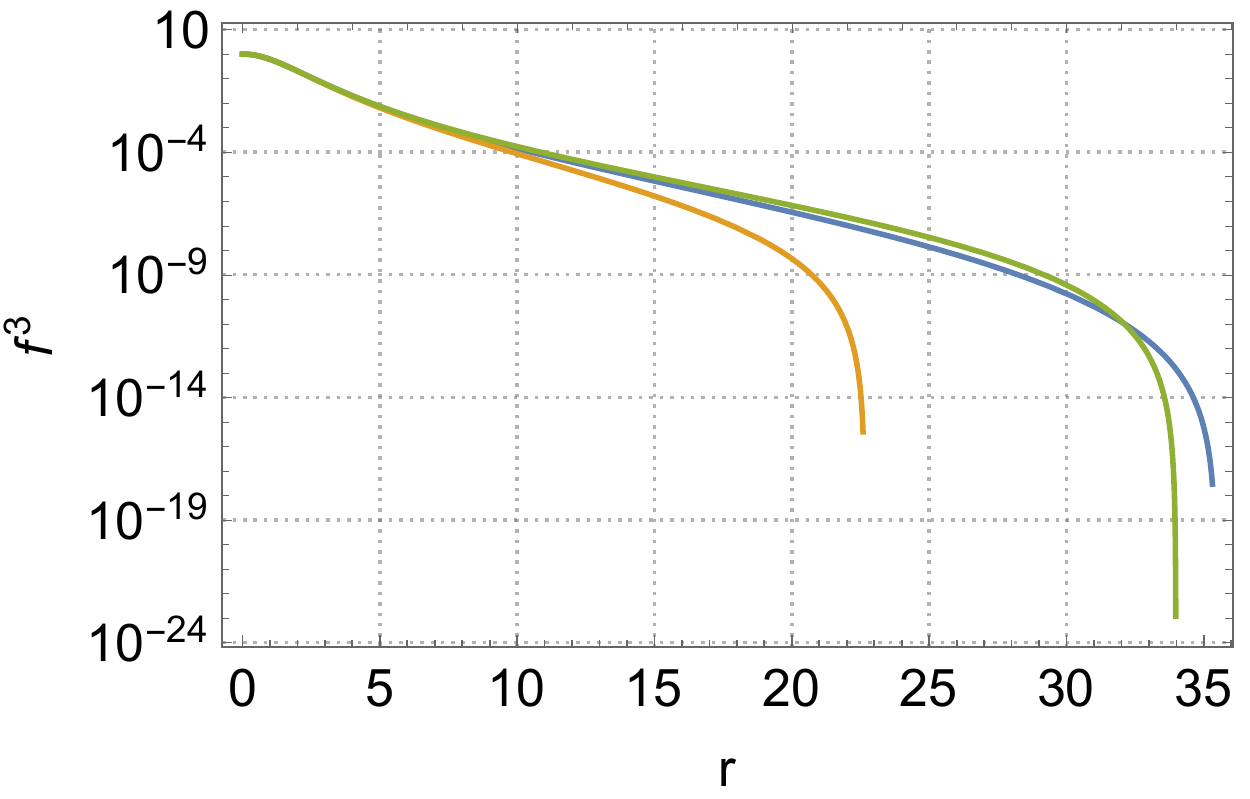}\includegraphics[width=3.0in]{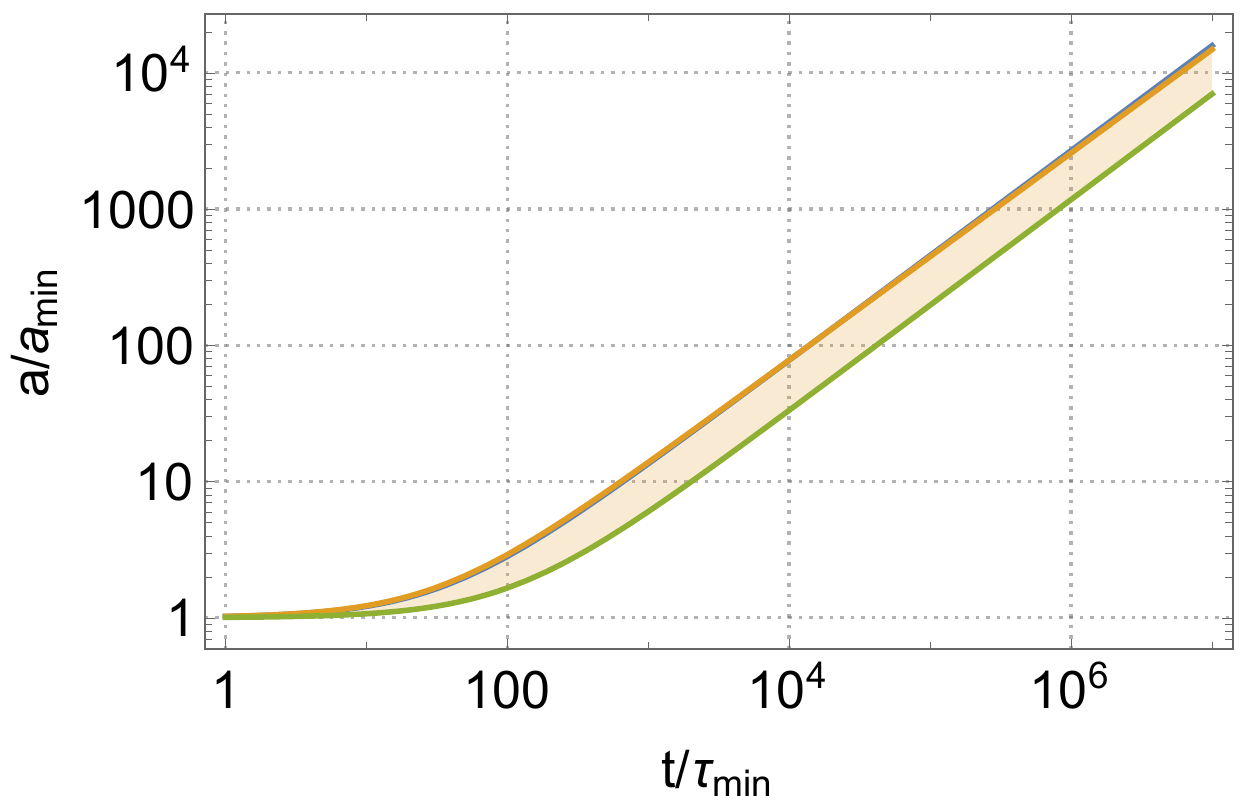}
\caption{Colours on line for three eigenvalues (\ref{para}). We show solutions to the eigenvalue problem (\ref{at}) and
(\ref{e-profile}). The homologous density profile $f^3$ is plotted
(left)as a function of dimensionless radius $r=R/a$ (\ref{a0}).
The density profile $f^3$ does not change in time, and it vanishes 
at the core outer radius $r=r_s$ (\ref{rs}). The scale function $a(t)/a_{\rm min}$ is plotted as in terms of $t/\tau_{\rm min}$, using units $a_{\rm min}$ (\ref{amin}) and $\tau_{\rm min}$ (\ref{taumin}). 
Note that $t\sim 10^7\tau_{\rm min}\sim {\mathcal O}(10)$ sec., 
depending on $\rho^{\rm max}_c$ value. 
}\label{fa}
\end{figure}

The theoretical scheme turns out to be an eigenvalue problem. 
All physical solutions  
of eigenvalues $(\gamma,\lambda_m)$ and $1<\gamma < 4/3$ have to be numerically found, 
consistently with physical conditions and observations. 
Using numerical approach (mathematics), we first 
reproduce the $\gamma=4/3$ solution \cite{Goldreich1980}, as a check of numerical argorithm.  
Furthermore, as shown in the left of Fig.~\ref{fa}, we find three physical solutions of homologous density profile $f^3$ 
corresponding to the selected eigenvalues $(\gamma,\lambda_m)$:
\begin{eqnarray}
\textcolor{orange!90!}{Orange}~(1.24,1.0\times 10^{-4});~ 
\blue{Blue}~(1.23,8\times 10^{-5});~
\green{Green}~(1.225,8\times 10^{-6}),
\label{para}
\end{eqnarray}
for which the homologous profile $f(r)$ becomes tangent to $f=0$ at the outer radius $r_s$, namely  
$f'(r_s)\approx 0$. 
The corresponding outer radius $r_s$ are:
\begin{eqnarray}
\textcolor{orange!90!}{Orange}~(r_s\approx 23);~ 
\blue{Blue}~(r_s\approx 36);~
\green{Green}~(r_s\approx 34),
\label{rs}
\end{eqnarray}
corresponding to three cases of eigenvalues $(\gamma,\lambda_m)$ (\ref{para}). In this article, we illustrate  
physical results by using three colour lines ($\textcolor{orange!90!}{orange},\blue{blue},\green{green}$),  
representing typically selected eigenvalues (\ref{para}). The shadow in between these colour lines indicates other possible physical solutions. 

These limiting eigenvalues $\lambda_m$ (\ref{para}) 
for $f'(r_s)\approx 0$ 
are reached when the core surface at $r_s$ (\ref{rs}) is in free fall, see discussions in Ref.~\cite{Goldreich1980}. 
Using Eq.~(\ref{e-profile}) and outer boundary condition 
$f'(r_s)\approx 0$, the mean core density $\bar\rho$ is 
obtained in terms of the core centre density 
\begin{eqnarray}
\bar\rho\equiv \frac{M}{(4\pi r^3_s/3)} = \frac{\int_0^{r_s} 4\pi r^2dr \rho }{(4\pi r^3_s/3)}
=\frac{\gamma}{4(\gamma-1)}\lambda_m \rho_c.
\label{meand}
\end{eqnarray}
The $M$ is the total mass of the homologously collapsing core. Both the core mean density $\bar\rho$ and the core centre density $\rho_c$ are functions of the time $t$. 

Given the eigenvalue $\lambda$, the first and second time 
integrations of the temporal equation (\ref{at}) give
\begin{eqnarray}
  \frac{1}{2}\dot{a}^2 &=& \frac{\lambda}{6}\frac{\gamma(2-\gamma)}{(\gamma-1)^2}\left(\frac{\kappa^{\frac{1}{\gamma-1}}}{\pi G}\right)^{\frac{\gamma-1}{2-\gamma}}a^{2\frac{1-\gamma}{2-\gamma}} +C,
	\label{a'}\\
	\frac{a(t)}{a_{\rm min}} &=& \left\{
1+ \Big[\frac{\lambda\gamma}{3(2-\gamma)(\gamma-1)^2}\Big]^{\frac{1}{2}}\Big(\frac{t}{\tau_{\rm min}}\Big)\right\}^{2-\gamma}.
		\label{af}
\end{eqnarray}
The basic gravitational time scale in homologous collapse is:
\begin{eqnarray}
\tau_{\rm min} &\equiv& \Big(\frac{1}{\pi G \rho_c^{\rm max}}\Big)^{1/2}=1.02\times 10^{-5}(\frac{\rho_c^{\rm max}}{\rho_0})^{-1/2}  ({\rm sec}),
\label{taumin}
\end{eqnarray}
and $\tau_{\rm min}=(3\gamma)^{1/2}a_{\rm min}$.
When the time approaches to the ending of homologous collapse, $t\rightarrow \tau_{\rm min}$, the scale function $a(t)$ approaches to 
the ``sound horizon'', $a(t)\rightarrow a_{\rm min}$, and core centre density approaches to its maximal value, $\rho(t)\rightarrow \rho^{\rm max}_c$ (\ref{amini}). As a comparison, the time scale $\tau_{\rm min}$ is larger than the free-fall time scale $\tau_{ff}=[3\pi/(32 G\bar\rho)]^{1/2}$ for $\rho_c>\bar\rho$  (\ref{meand}).
In fact the time scale $\tau_{\rm min}$  
characterises the macroscopic time scale $\tau_{\rm grav}$ that we discussed in the time scale hierarchy (\ref{Thy}). 

The constant of integration $C$ in Eq.~(\ref{a'}) determines 
the initial velocity of collapse. Its value has no great effect on the solution $a(t)$, when the scale function $a(t)$ approaches to 
the ``sound horizon'' $ a_{\rm min}$ (\ref{amini}). We put $C=0$ and 
$\lambda>0$ to obtain the second integration (\ref{af}).
The gravitational collapse process goes in the inverse time direction from the starting time $t>\tau_{\rm min}$ when $a=a(t)$ and $\rho_c=\rho_c(t)$, to the ending time $\tau_{\rm min}$  
when $a\rightarrow a_{\rm min}$ and $\rho_c(\tau_{\rm min})\rightarrow 
\rho^{\rm max}_c$. Namely, when the core starts homologous collapse at time $t$, its centre density is $\rho_c=\rho_c(t)$ and centre sound velocity $v_s^c=v_s^c(t)$, given the core mass $M$ and radius 
$r_s$ (\ref{meand}).
 
\begin{figure}
\includegraphics[width=3.0in]{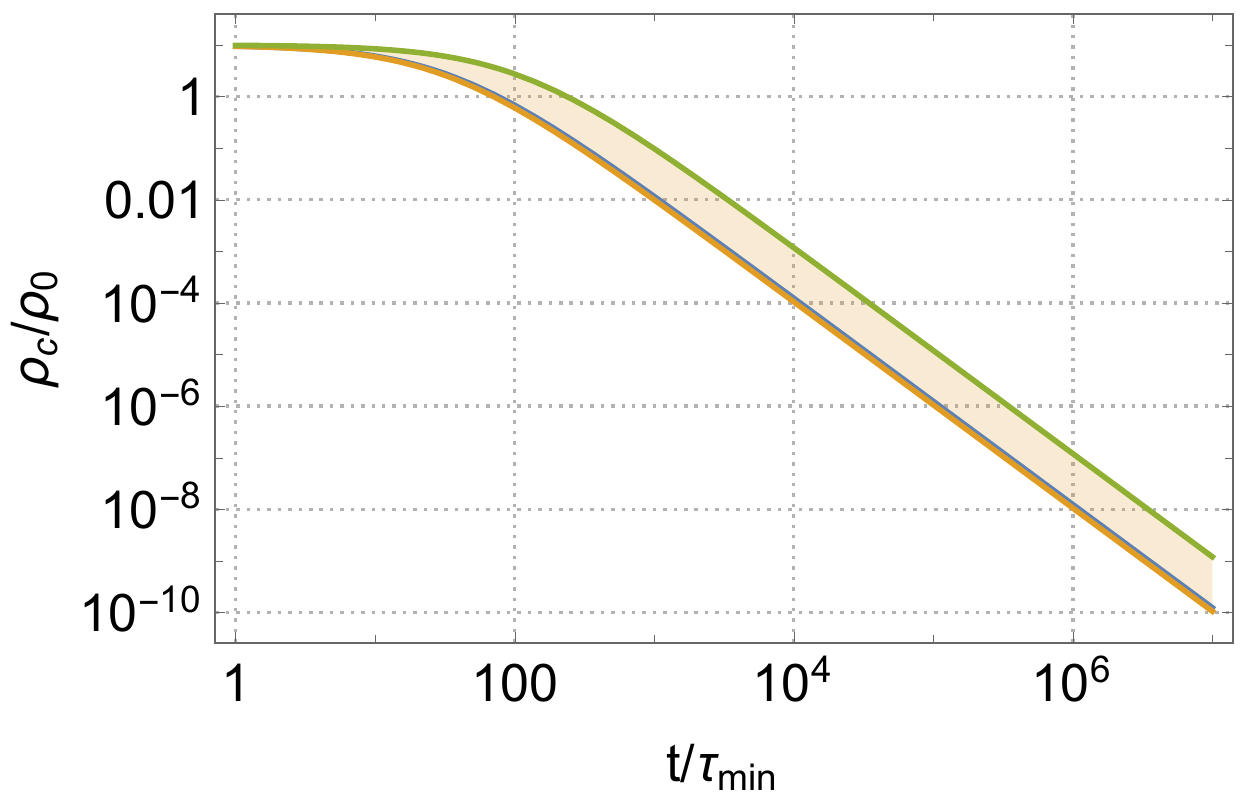}\includegraphics[width=3.0in]{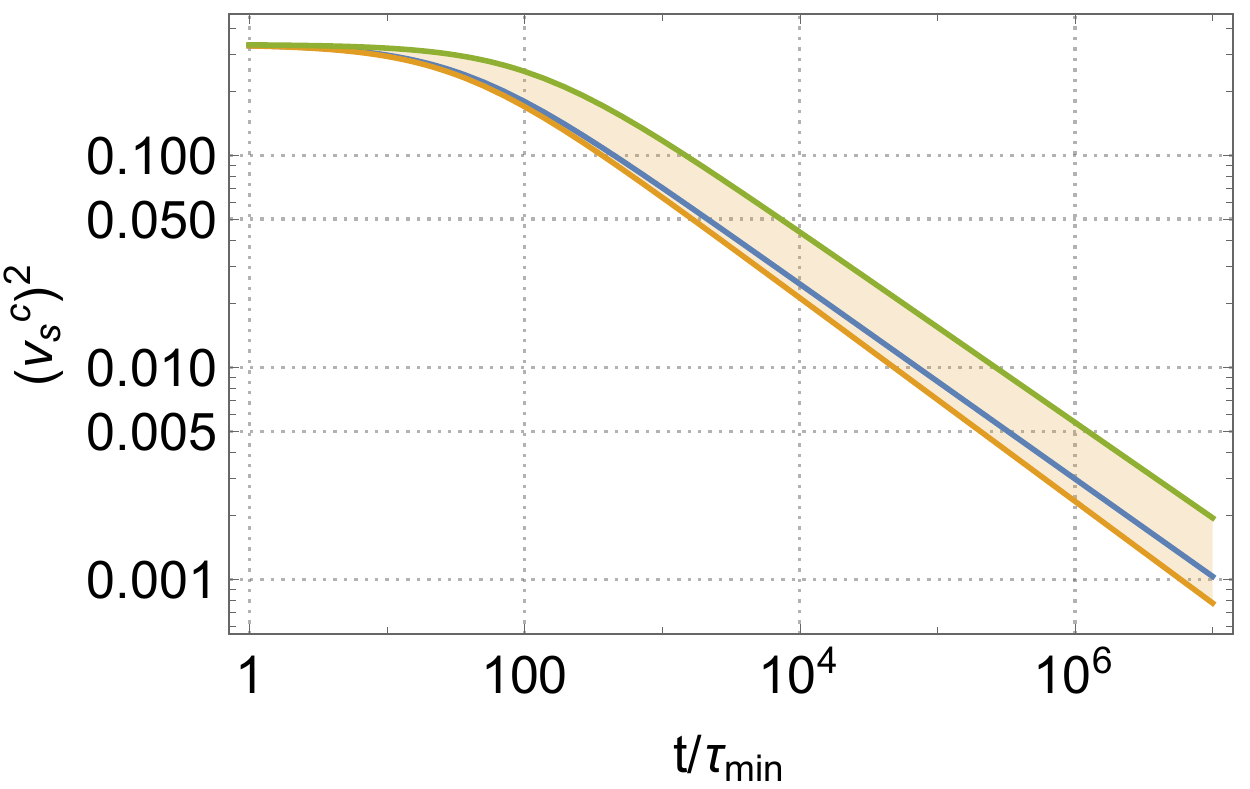}
\caption{Colours on line for three eigenvalues (\ref{para}). The core centre density $\rho_c$ (left) and sound velocity $(v_s^c)^2$ (right) are plotted as functions of time in an homologous collapsing process. Their final values end up to $t=\tau_{\rm min}$ when $\rho_c=\rho^{\rm max}_c=10\rho_0$ and $(v_s^c)^2=1/3$. Their initial values $\rho_c(t)$ and $(v_s^c)^2(t)$depend on the initial time $t$, given the core mass $M$ and radius $r_s$ (\ref{meand}) and the validity of homologous collapse description. 
For instance, $t=10^7~ \tau_{\rm min}\sim {\mathcal O}(10)$ second, 
when $\rho_c= 10^{-(11\sim 10)}\rho^{\rm max}_c= 10^{-(10\sim 9)}\rho_0$ and $(v_s^c)^2=(0.001 \sim 0.003)$. 
}\label{vrho}
\end{figure}

As results, the scale function 
$a(t)$ (\ref{af}) is shown in the right of Fig.~\ref{fa}. The density $\rho_c$ and sound velocity $v_s^c$ at the core centre are shown in Fig.~\ref{vrho}. These figures show the variations of scale function $a(t)$, centre density $\rho_c(t)$ and sound velocity $v_s^c(t)$ in homologous collapsing process from $t\sim 10$ seconds to  $t\sim 10^{-5}$ seconds. The centre scale function $a(t)$ (\ref{a0}) and sound velocity $(v_s^c)^2(t)$ (\ref{sound}) are in terms of the centre density 
$\rho_c(t)$. All physical quantities are then functions 
of $\rho_c(t)$. While, at the time $t$ when the core starts homologous collapse, the initial value $\rho_c(t)$ depends on the baryon core total mass $M$ and its homologously collapsing radius $r_s$ (\ref{meand}).

\begin{figure}
\includegraphics[width=3.0in]{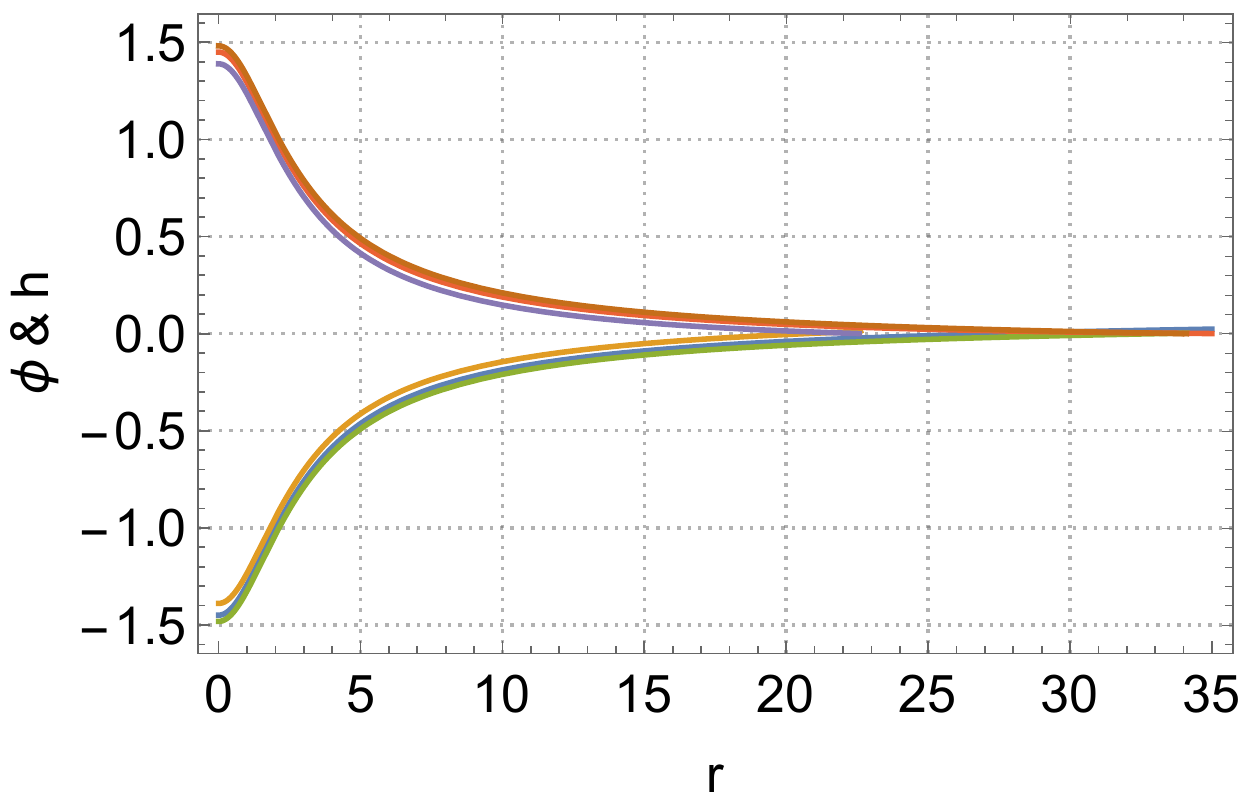}\includegraphics[width=3.0in]{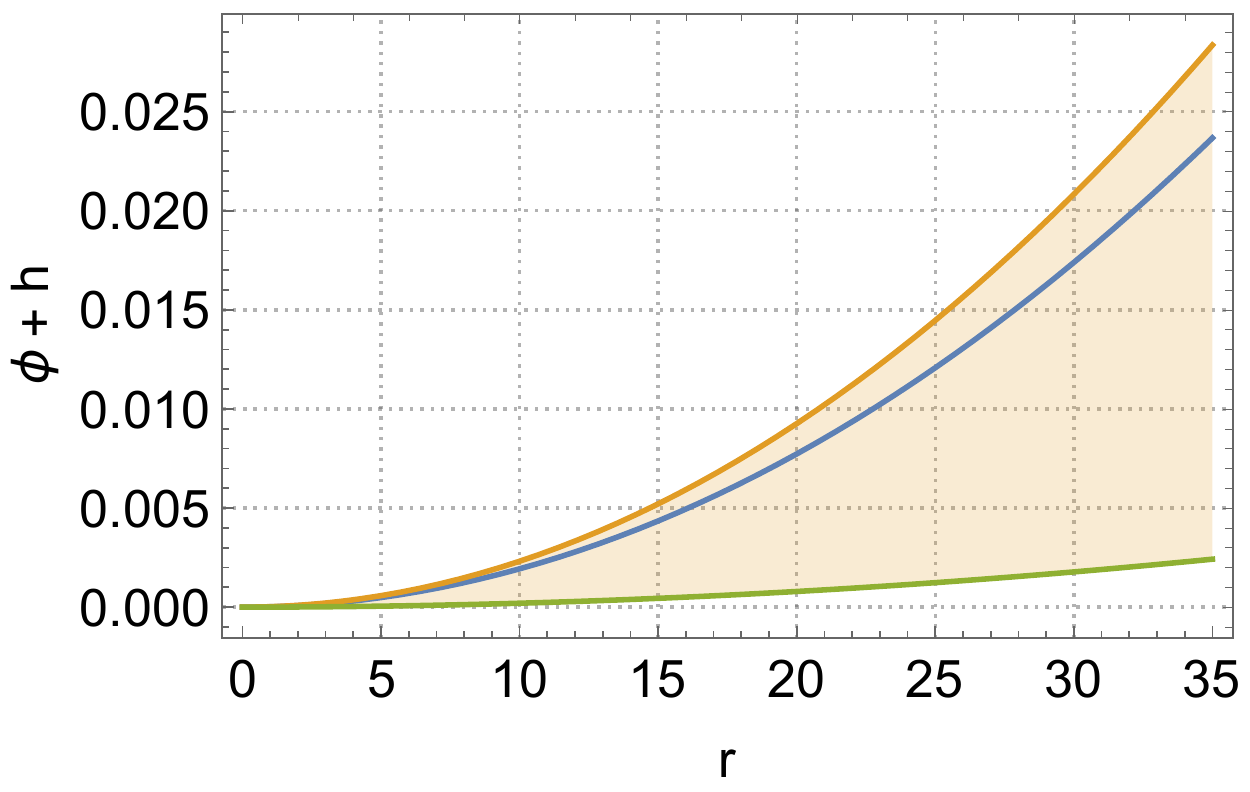}
\caption{Colours on line for three eigenvalues (\ref{para}). Using solutions (\ref{hphi}) for maximal sound velocity $(v_s^c)^2=1/3$, the positive 
heat function $h$, negative gravitational potential 
$\phi$ (left) and their sum (right) are plotted 
as a function of dimensionless radius $r$. These homologous 
function ($h,\phi$) profiles 
do not depend on the time $t$. However their amplitudes depend on the sound velocity $(v_s^c)^2(t)$. The time variations of these homologous 
profile amplitudes can be obtained by multiplying the sound velocity given by the right of Fig.~\ref{vrho}.
}\label{fhphi}
\end{figure}

To illustrate the gravitational and thermal dynamics 
of homologous collapse, we also plot in 
Fig.~\ref{fhphi} the positive heat function (\ref{heat}) and negative  gravitational potential (\ref{phi1})
\begin{equation}
h=\frac{(v_s^c)^2}{\gamma-1}f^{3(\gamma-1)};\quad
\phi=\frac{(v_s^c)^2}{\gamma-1}\Big(\frac{\lambda}{6} r^2 - f^{3(\gamma-1)}\Big)
\label{hphi}
\end{equation}
and their sum $h+\phi=\frac{(v_s^c)^2}{\gamma-1}\frac{\lambda}{6} r^2$ in the Euler equation (\ref{euler1T}). The results depend on the thermal index $\gamma$ in the equation of state (\ref{eos}) 
and the eigenvalue $\lambda >0$. The latter
relates to the material binding energy of baryon fluid per energy density 
$\rho$ \cite{Goldreich1980}. Indeed, our solutions (\ref{para}) and (\ref{rs}) show
that the binding energy $\lambda_m$ value decreases, the outer boundary $r_s$ 
becomes larger, i.e., the core is less bound, and the thermal index $\gamma$ 
becomes smaller, i.e., the EoS (\ref{eos}) becomes soften.

\section{Photon-pair sphere formation and properties}\label{psphere} 

\subsection{Photon-pair sphere temperature and energy density}

\begin{figure}[t]
\includegraphics[width=3.0in]{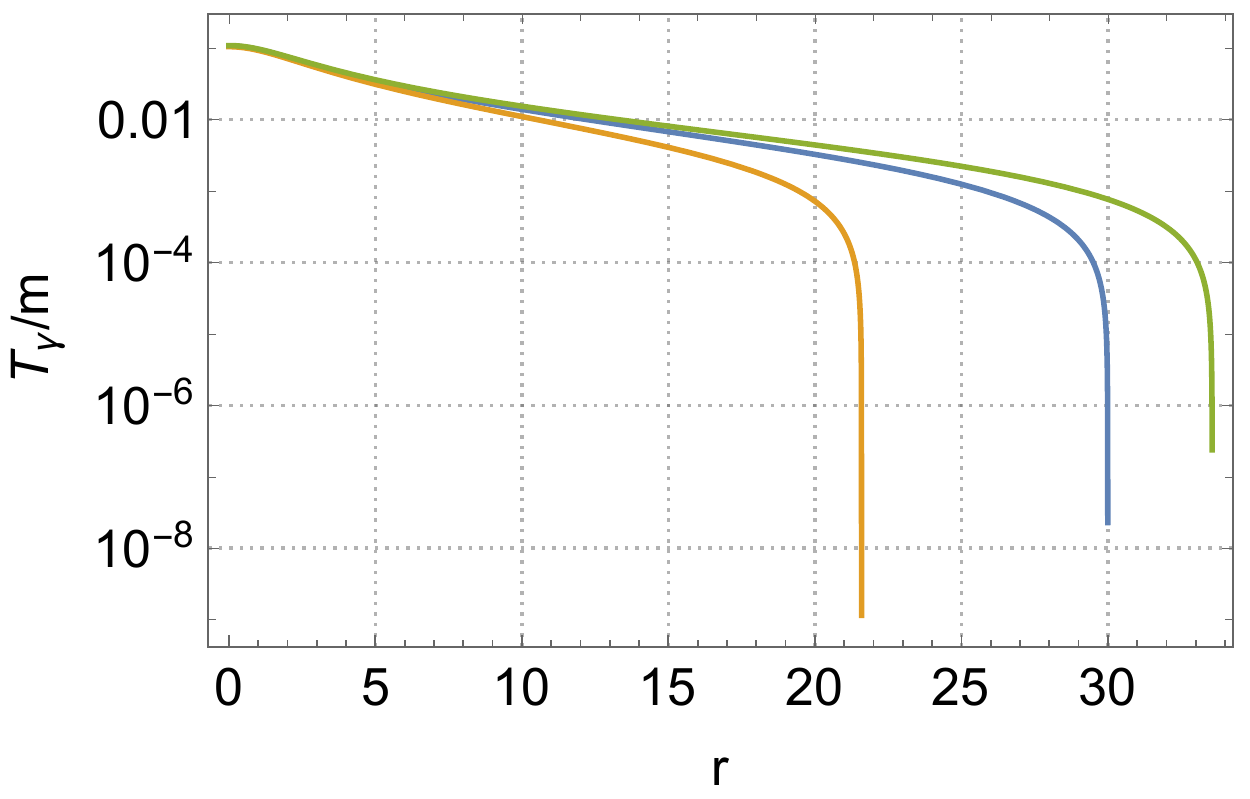}\includegraphics[width=3.0in]{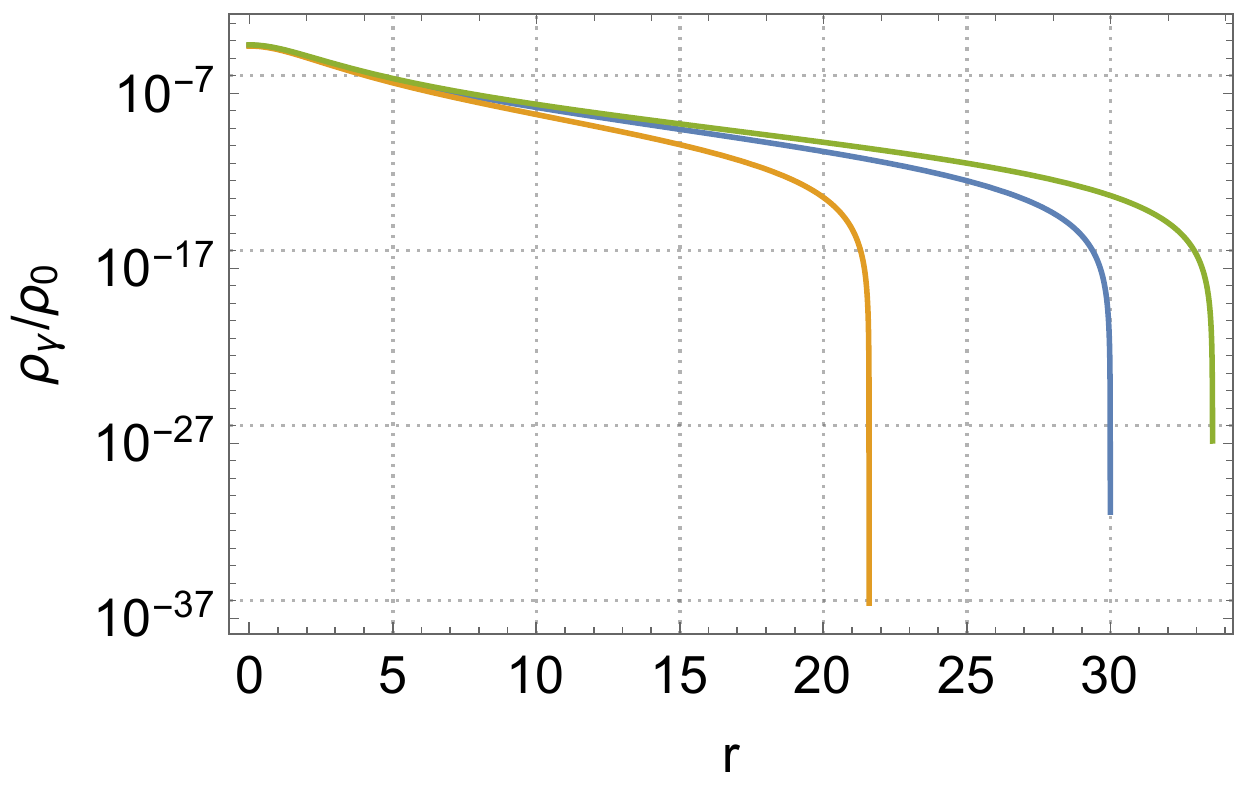}
\caption{Colours on line for three eigenvalues (\ref{para}). Using results (\ref{dgravT1}) 
and (\ref{gammaEd}) for maximal sound velocity $(v_s^c)^2=1/3$, we plot the homologous profiles of temperature $T_\gamma/m$ (left) and energy density $\rho_\gamma/\rho_0$ (right) 
as a function of dimensionless radius $r$. Both profiles vanish at the photon-pair sphere radius $r_\gamma$ (\ref{para1}), which is smaller than the core outer radius $r_s$(\ref{rs}), namely $r_\gamma < r_s$, see left of Fig.~\ref{fa}. Note that the typical hadron mass $m\approx 1$ GeV and 
nuclear saturation density $\rho_0\approx 2.4\times 10^{35} {\rm ergs}/{\rm cm}^3$. These homologous profiles do not depend on the time $t$. But their absolute values depend on time-dependent sound 
velocity $(v_s^c)^2(t)$, see Eqs.~(\ref{dgravT1}) and (\ref{gammaEd}).
The time evolutions of photon-pair sphere temperature $T_\gamma(r,t)/m$ (\ref{dgravT1}) and energy density  $\rho_\gamma(r,t)/\rho_0$ (\ref{gammaEd}) are given in Fig.~\ref{timev} in Appendix \ref{app}.
}\label{ted}
\end{figure}

On the basis of {\it adiabatic} approximation discussed below the time 
scale hierarchy (\ref{Thy}), we use the ``{\it local}'' virial theorem
(\ref{dgravP}) or (\ref{dgravT}) 
to obtain the baryon temperature $T$ (\ref{dgravT}), 
then the photon-pair sphere temperature $T_\gamma$ (\ref{Tphoton})  
\begin{eqnarray}
\frac{T_\gamma}{m} \approx 0.21 (v^c_s)^2{\mathcal T}(r),\quad 
{\mathcal T}(r)\equiv \frac{1}{6\gamma (\gamma -1)}
\left[(7\gamma -6)f^{3(\gamma-1)} - \frac{\lambda}{6}\gamma r^2 \right].
\label{dgravT1}
\end{eqnarray}
The time-dependent centre sound velocity $(v^c_s)^2$ comes from Eq.~(\ref{sound}). 
The photon-pair sphere temperature 
$T_\gamma$ reaches its maximum 
\begin{eqnarray}
\frac{T_\gamma^{\rm max}}{m} \approx 0.21  (v^c_s)^2 
\frac{(7\gamma -6)}{6\gamma (\gamma -1)},
\label{tmax}
\end{eqnarray} 
at 
photon-pair sphere centre $r=0$.  The photon-pair sphere temperature has no any lower limit, 
and it vanishes at the outer boundary $r=r_\gamma <r_s$, determined 
by ${\mathcal T}(r_\gamma)=0$. 

Furthermore, we use 
photon productions (\ref{gammaN}) 
and (\ref{rhog}) by baryon collisions to approximately obtain 
the  photon-pair sphere energy and number densities
\begin{eqnarray}
\frac{\rho_\gamma}{\rho_0}&\approx & 1.29\times 
\frac{4\alpha\alpha_s}{3\pi}\Big(\frac{T_\gamma}{m}\Big)^4\ln\left( 1+ \frac{2.9}{4\pi\alpha_s}\right),
\label{gammaEd}\\
\frac{n_\gamma}{n_0} &\approx & 6.16 \times 
\frac{4\alpha\alpha_s}{3\pi}\Big(\frac{T_\gamma}{m}\Big)^3\ln\left( 1+ \frac{2.9}{4\pi\alpha_s}\right).
\label{gammaNd}
\end{eqnarray}
In Fig.~\ref{ted}, we plot the photon-pair sphere temperature $T_\gamma/m$ 
and energy density $\rho_\gamma/\rho_0$ for $(v^c_s)^2=1/3$. 
As a self-consistency check, 
we find that the baryon temperature $T$ and photon temperature $T_\gamma$ 
are much smaller than baryon mass 
($T_\gamma \approx 0.21 T \ll m$). The condition is fully filled 
for applying the virial theorem (\ref{dgravP}). 

In addition, the photon energy density $\rho_\gamma$ (\ref{rhog}) is much smaller 
than not only baryon mass-energy density $\rho$, but also baryon heat energy 
density $(3/2)T\rho/m$ (\ref{dgravP}). This justifies {\it a posterior} that 
the photon-pair sphere energy density $\rho_\gamma$ and heat function (thermal potential) 
$h_\gamma=\rho_\gamma+p_\gamma$ can be neglected in homologous collapse dynamics, 
studied in previous Sec.~\ref{sec:1}. Further discussions 
on the effective photon heat function will be presented in Sec.~\ref{form}. 

The ``instant'' and ``locality'' of microscopic processes discussed below the 
time hierarchy (\ref{Thy}) are necessary conditions that, and reasons 
why photons produced can form an optic thick sphere or jet at macroscopic scales. 
Such trapped photons are 
in collective motions through density perturbations 
and/or other hydrodynamical processes of time scale $\sim |{\bf R}|/v_s$.

\subsection{Photon-pair sphere size and opacity} 

\begin{figure}
\includegraphics[width=3.0in]{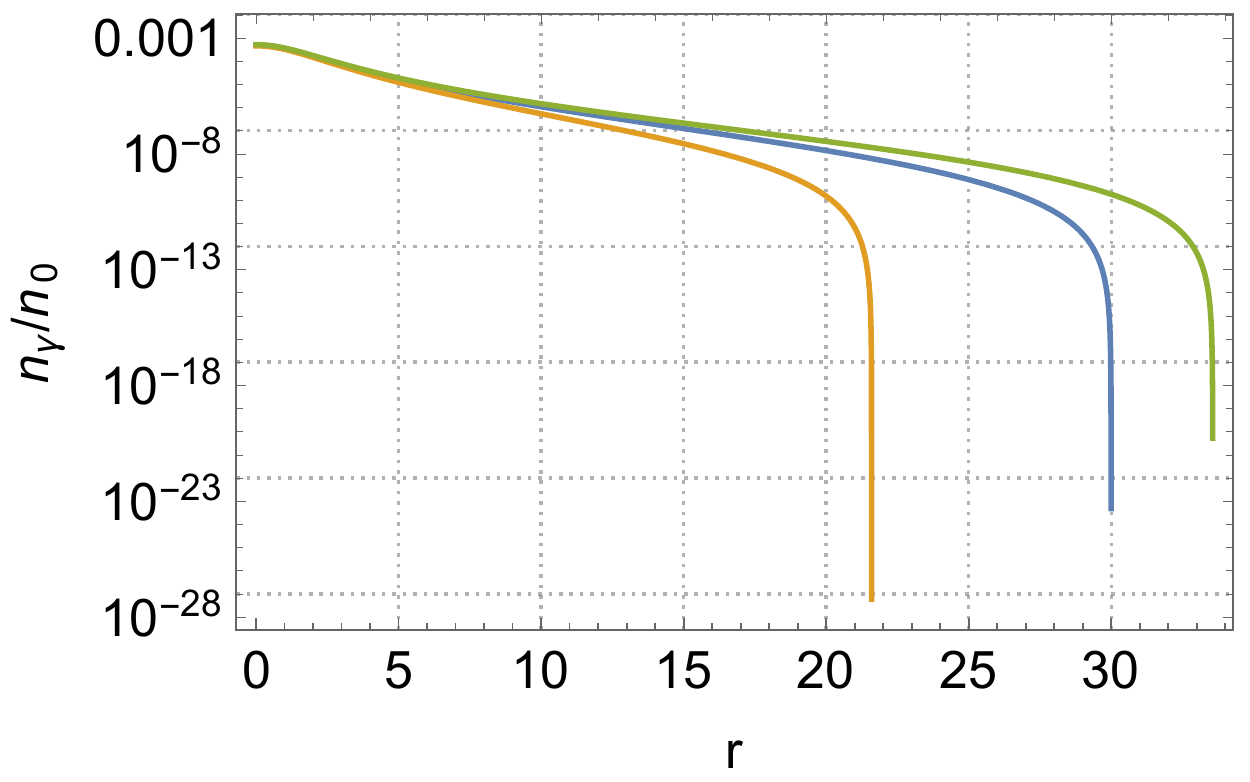}
\includegraphics[width=3.0in]{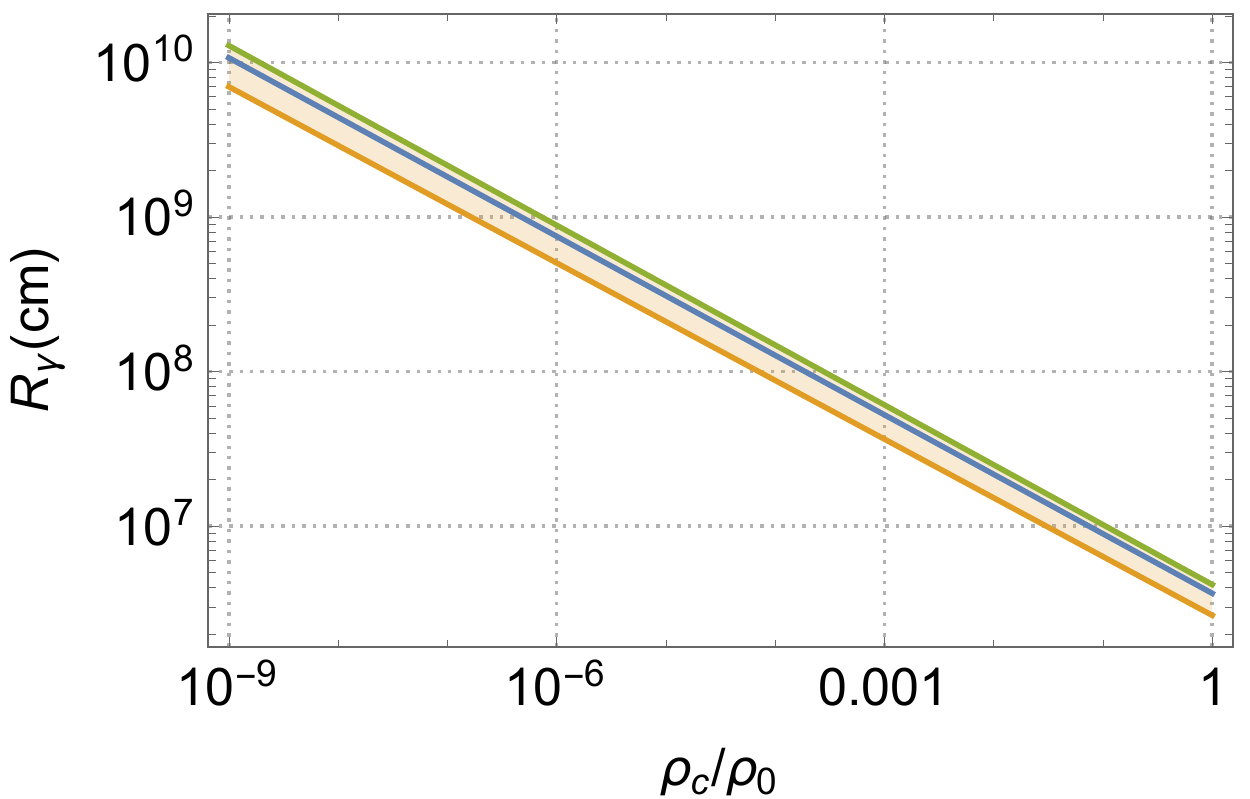}
\caption{Colours on line for three eigenvalues (\ref{para}). 
For maximal sound velocity $(v_s^c)^2=1/3$, we plot (left)
the photon number density 
$n_\gamma/n_0$ (\ref{gammaNd}) vanishes at outer boundary $r_\gamma$ (\ref{para1}). The 
photon-pair sphere size $R_\gamma=ar_\gamma$ (\ref{totn}) (right) is 
plotted as a function of the initial core centre density $\rho_c(t)$ at the time $t$ 
when core homologous collapse starts. 
The nuclear saturation density is
 $n_0\approx 1.6\times 10^{38}/{\rm cm}^3$ or
 $\rho_0\approx 2.4\times 10^{35} {\rm ergs}/{\rm cm}^3$. 
The time evolution of photon-pair sphere number density $n_\gamma(r,t)/n_0$ (\ref{gammaNd}) profile is given in Fig.~\ref{timev} in Appendix \ref{app}.}\label{ndensize}
\end{figure}

As shown in Fig.~\ref{ted}, the photon-pair sphere temperature and energy density profiles 
vanish at diemensionless radius $r_\gamma<r_s$:
\begin{eqnarray}
\textcolor{orange!90!}{Orange}~(r_\gamma=21.59);~ 
\blue{Blue}~(r_\gamma=30.00);~
\green{Green}~(r_\gamma=33.56),
\label{para1}
\end{eqnarray}
corresponding to three cases of eigenvalues $(\gamma,\lambda_m)$ (\ref{para}) 
and $r_s$ (\ref{rs}). 
This determines the photon-pair sphere radial size $R_\gamma=a(t)r_\gamma$ (\ref{a0}). Using Eqs.~(\ref{soundc}), (\ref{amin}) and (\ref{af}), we obtain
\begin{eqnarray}
R_\gamma &=& 3.06\times 10^{5}(3\gamma)^{-1/2}  r_\gamma \left(\frac{\rho_0}{\rho_c^{\rm max}}\right)^{\frac{\gamma-1}{2}} \left(\frac{\rho_0}{\rho_c}\right)^{\frac{2-\gamma}{2}} ({\rm cm}),
\label{Rsize}
\end{eqnarray}
which is plotted in Fig.~\ref{ndensize} as a function of initial core centre density $\rho_c=\rho_c(t)$, when a homologous collapse initiates. This result shows that 
the photon-pair sphere radius $R_\gamma$ depends weakly on the eigenvalue 
$r_\gamma(\gamma,\lambda_m)$, which is independent of the time $t$. 
This behaviour is shown in some details by the homologous profiles given 
in Fig.~\ref{timev} in Appendix \ref{app}. The photon-pair sphere radius $R_\gamma$ 
decreases, as the initial core centre density $\rho_c(t)$ increases. 
The reason will be given below when we discuss the total photon-pair sphere energy and number.

Moreover, we show in Fig.~\ref{ndensize} the photon-pair sphere 
number density $n_\gamma$ (\ref{gammaNd}) 
and photon-pair sphere size $R_\gamma=a r_\gamma$ (\ref{Rsize}). The results show that 
the photon number density $n_\gamma$ is in the range of $(10^{37}\sim 10^{30})/{\rm cm}^3$,
and photon-pair sphere size $R_\gamma$ is in the range of $(10^{7}\sim 10^{10})~{\rm cm}$. The
photon-pair sphere is deeply opaque, since the photon mean-free path 
$\lambda_\gamma =(\sigma_\gamma n_\gamma)^{-1}$ 
is much smaller than the photon-pair sphere size $R_\gamma$, 
namely $\sigma_\gamma n_\gamma R_\gamma\gg 1$. 

On the other hand, we find in Fig.~\ref{ted} that the photon-pair sphere 
temperature $T_\gamma$ can be well above the critical energy threshold 
$2m_e=1.02$ MeV of electron-positron ($e^+e^-$) pair production. 
The photon-pair sphere energy density $\rho_\gamma$ can be well above the 
critical energy density $\rho_e=m_e^4=5.93\times 10^{-11}\rho_0$ 
of electron-positron pairs.
This means that the photons 
are quickly thermalised to form an electro-position-photon plasma characterised by the thermalisation
time scale $\tau_{\rm therm}$ (\ref{Tgamma}) \cite{Preparata1998,Ruffini2010,Ruffini2003a}, as briefly  
discussed in Sec.~\ref{virial}.

\begin{figure}
\includegraphics[width=3.0in]{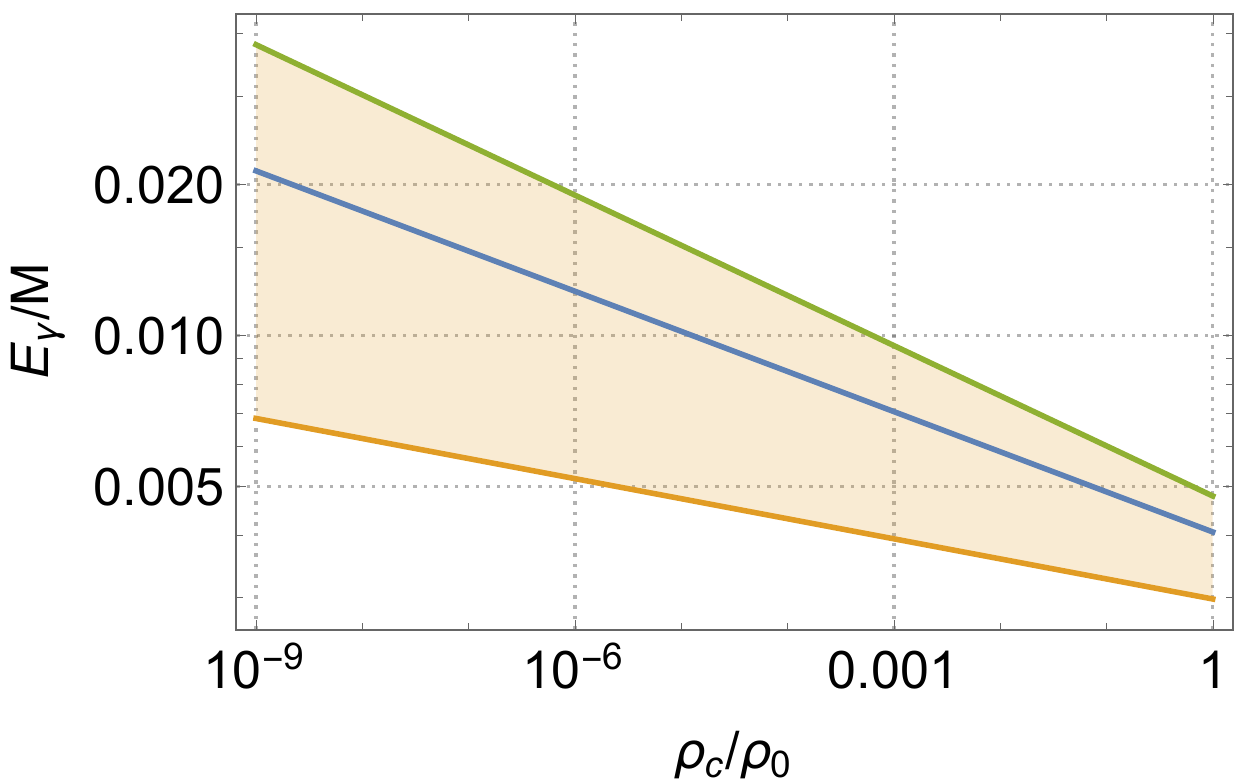}
\includegraphics[width=2.8in]{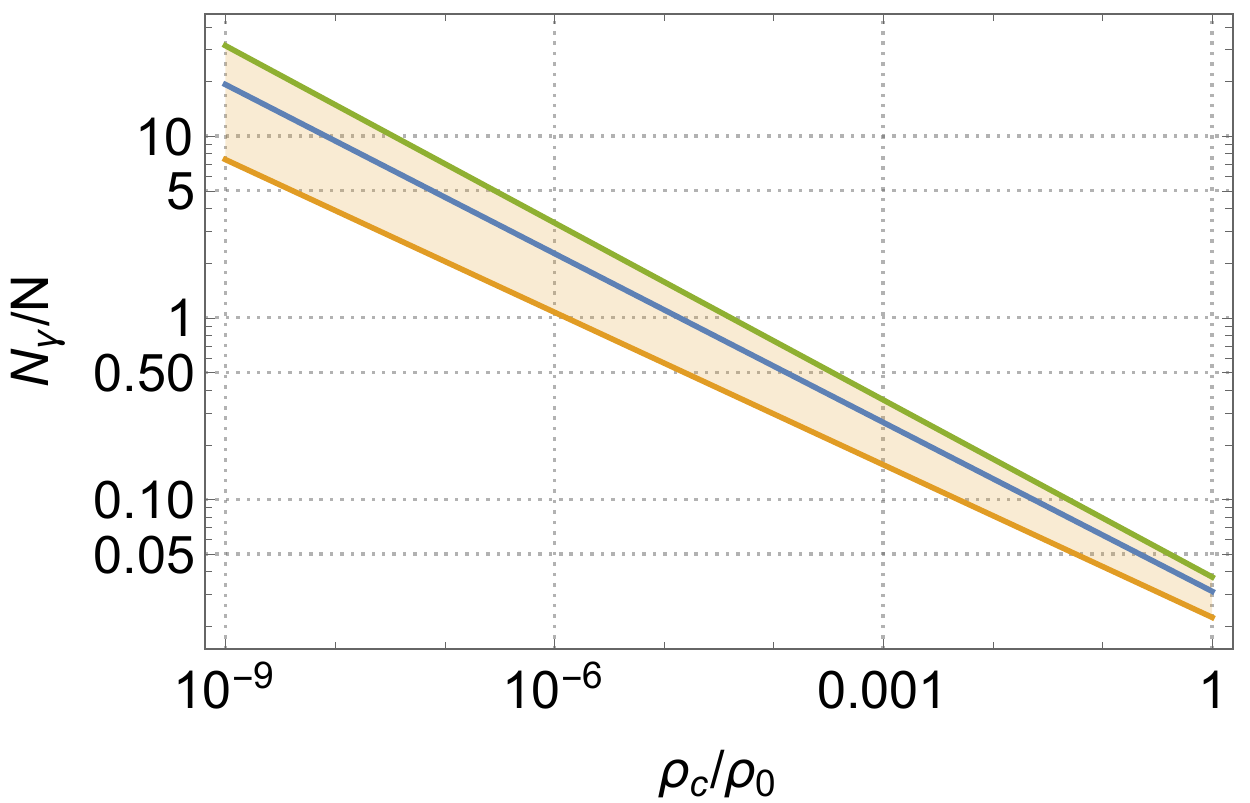}
\includegraphics[width=3.0in]{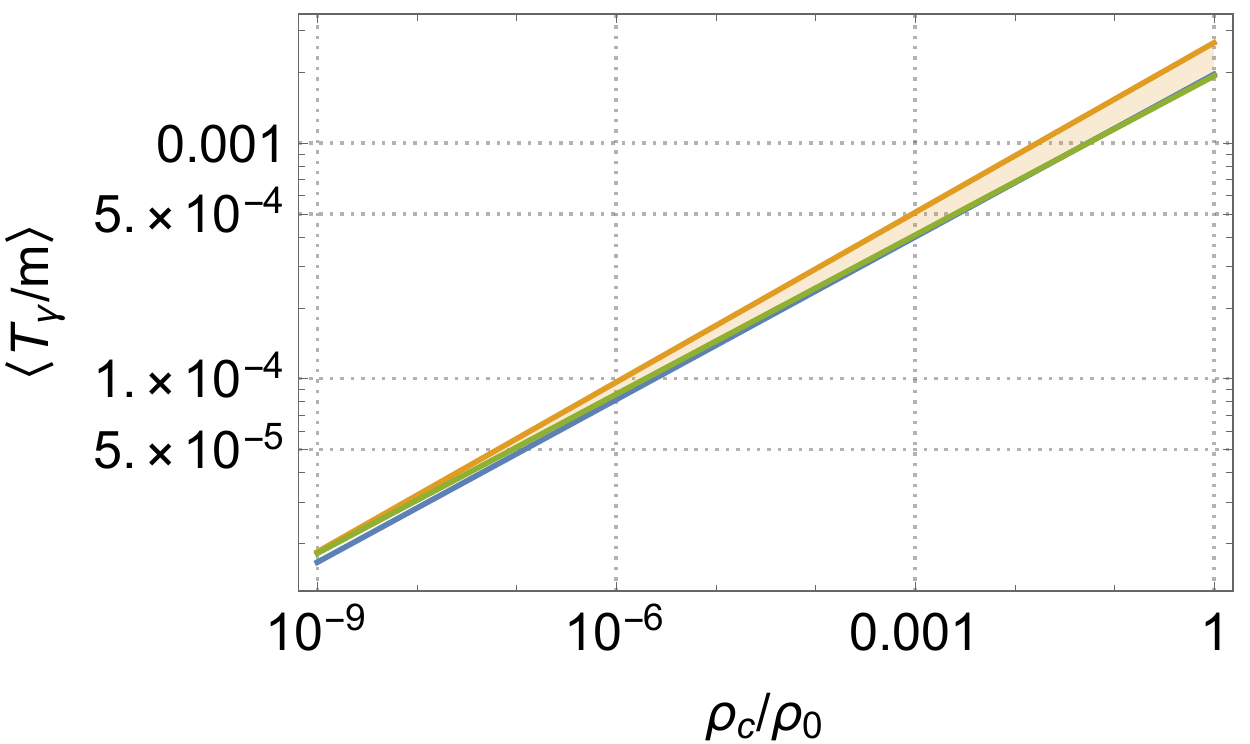}
~~~~~\includegraphics[width=2.9in]{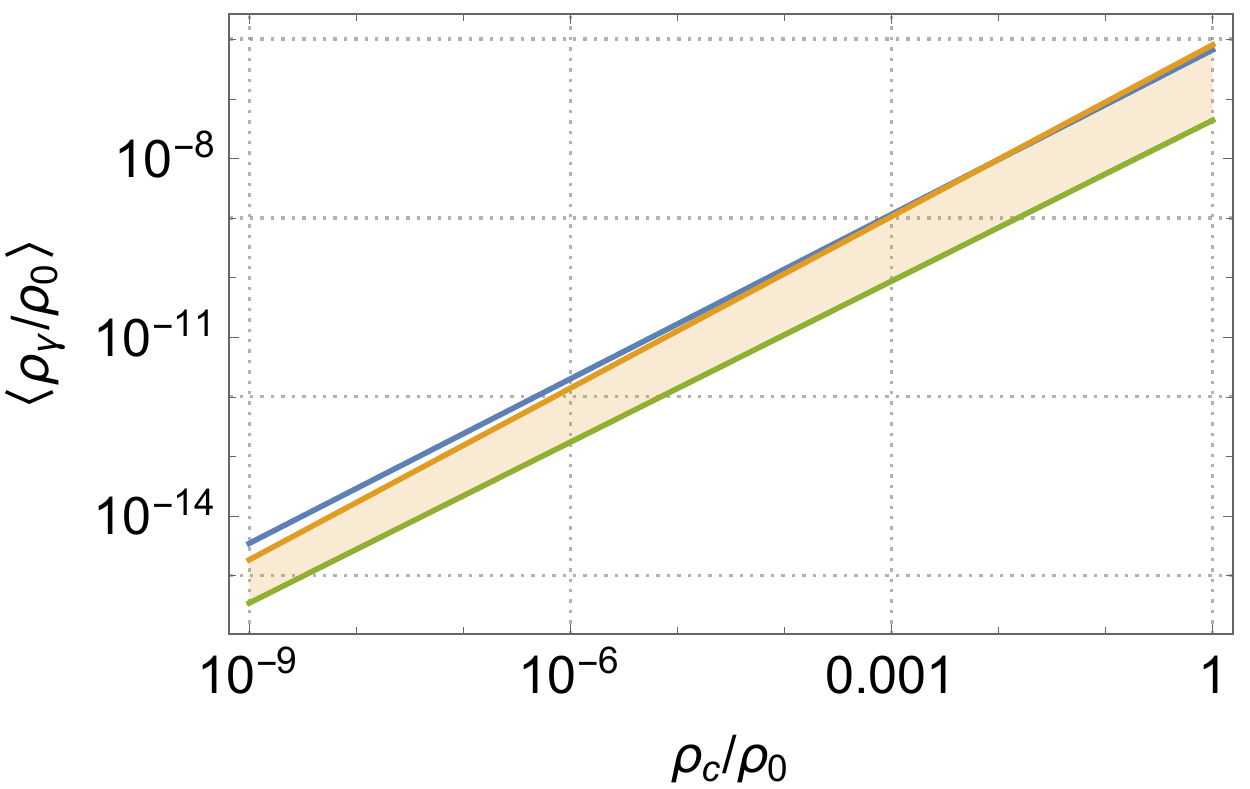}
\caption{Colours on line for three eigenvalues (\ref{para}). Top panel: the total photon
sphere energy $E_\gamma/M$ (\ref{tote}) (left) 
and number $N_\gamma/N$ (\ref{totn}) (right); bottom panel: the mean photon-pair sphere
temperature $\langle T_\gamma\rangle/m$ (\ref{atem}) (left) 
and energy density $\langle \rho_\gamma\rangle /\rho_0$ (\ref{aeden}) (right),  
are plotted as a function of the initial core 
centre density $\rho_c$, when homologous collapse starts.  
The nuclear saturation density $\rho_0\approx 2.4\times 10^{35} {\rm ergs}/{\rm cm}^3$
and typical hadron mass $m\approx 1$ GeV. $M$ and $N$ are total mass and baryon 
number of the homologous collapsing core, 
$M_\odot\approx 1.8\times 10^{54}$ergs and $N_\odot\approx 1.2\times 10^{57}$.
The ratio $\rho_c(t)/\rho_0$ relates to the time $t/\tau_{\rm min}$ when the collapse 
start, see Fig.~\ref{vrho}.
}\label{toten}
\end{figure}

\subsection{Total photon-pair sphere energy and number}

The total relativistic particle energy $E_\gamma$ and number $N_\gamma$ of photon-pair sphere can be obtained by numerically integrating the energy density (\ref{gammaEd}) and number density (\ref{gammaNd}) over the photon-pair sphere volume $\int d^3R=a^3\int 4\pi r^2dr$. For comparison, we also compute the total baryon core 
 mass $M=\int \rho d^3R=\rho_c a^3F_0$ and number $N=M/m$, where $F_0\equiv \int_0^{r_s} 4\pi r^2 drf^3$. Defining coefficients $Y_n \equiv  
\int_0^{r_\gamma}4\pi r^2dr [{\mathcal T}(r)]^n$, we obtain 
\begin{eqnarray}
\frac{E_\gamma}{M}
&\approx & 1.29\times 
\left(\frac{1}{3}\right)^4 \frac{4\alpha\alpha_s}{3\pi}\ln\big(1+\frac{2.9}{4\pi\alpha_s}\big)\frac{Y_4}{F_0}\big(\frac{\rho_c^{\rm max}}{\rho_0}\big)^{4(1-\gamma)}
\big(\frac{\rho_0}{\rho_c}\big)^{5-4\gamma}
\label{tote}\\
\frac{N_\gamma}{N}
&\approx & 6.16\times 
\left(\frac{1}{3}\right)^3\frac{4\alpha\alpha_s}{3\pi}\ln\big(1+\frac{2.9}{4\pi\alpha_s}\big)\frac{Y_3}{F_0}\big(\frac{\rho_c^{\rm max}}{\rho_0}\big)^{3(1-\gamma)}
\big(\frac{\rho_0}{\rho_c}\big)^{4-3\gamma}.
\label{totn}
\end{eqnarray}
We expressed them in terms of the core centre density 
$\rho_c=\rho_c(t)$ at the time $t$ when the core starts to homologous collapse.  
Whereas, the time evolution of core centre density $\rho_c$ can be found 
in Fig.~\ref{vrho} or Eq.~(\ref{soundc}). 

Results (\ref{tote}), (\ref{tote}) and (\ref{totn}) show the photon-pair sphere size $R_\gamma$, 
energy $E_\gamma$ and number $N_\gamma$ depend on the initial core centre density 
$\rho_c$ and the averaged thermal index $\gamma$. 
When the massive core starts to homologous collapse,
the smaller initial value of the core centre density 
$\rho_c(t)$ is, the more extend core mass density profile $\rho(r,t)$ is, 
provided total core mass $M$ and baryon number $N$ fixed and conserved. 
As a consequence, the more gravitational energy can be gained in collapse 
and converted to the photon-pair sphere energy $E_\gamma$ so that 
the photon-pair sphere size $R_\gamma$ and number $N_\gamma$ are larger as well. 
However, in contrast, these quantities are also power functions of the averaged 
thermal index $\gamma$, 
that represents the core repulsive reaction against gravitational collapse and 
energy gain. As shown in Fig.~\ref{toten}, we notice that for a
large variation of the core centre density $\rho_c/\rho_0$, the total photon-pair sphere energy
$E_\gamma/M$ slowly varies within a order of magnitude. As an illustration, 
in Fig.~\ref{toten}, we make plots starting from the core centre density 
$\rho_c(t)\sim 10^{-9} \rho_0$ corresponds to the core collapsing time 
$t\approx 10^{7}\tau_{\rm min}\sim 10$ sec., see Fig.~\ref{vrho}. 
From these figures \ref{fa}, \ref{vrho},\ref{ndensize}, and \ref{toten}, 
people can find baryon core and photon-pair sphere properties, given an initial 
collapsing core density $\rho_c(t)/\rho_0$ and time $t/\tau_{\rm min}$, e.g., 
$\rho_c(t)/\rho_0\sim 10^{-3}$ and time $t/\tau_{\rm min}\sim 10^3$.  

In addition, we define the mean photon-pair sphere temperature and energy density as averaged 
values of $T_\gamma$ (\ref{dgravT1}) and $\rho_\gamma$ (\ref{gammaEd}) 
over photon-pair sphere volume $(4\pi/3)r_s^3$, 
\begin{eqnarray}
\frac{\langle T_\gamma\rangle}{m}&\equiv&0.21(v_s^c)^2 \frac{Y_1}{(4\pi/3)r_\gamma^3}= 0.21\left(\frac{1}{3}\right) \frac{Y_1}{(4\pi/3)r_\gamma^3}\left(\frac{\rho_0}{\rho_c^{\rm max}}\right)^{\gamma-1}\left(\frac{\rho_c}{\rho_0}\right)^{\gamma-1},
\label{atem}\\
\frac{\langle \rho_\gamma\rangle}{\rho_0}
&\equiv &  \frac{E_\gamma}{(4\pi/3)a^3r_\gamma^3\rho_0} =\left(\frac{E_\gamma\bar\rho}{M\rho_0}\right) \left(\frac{r_s}{r_\gamma}\right)^3=\frac{\gamma\lambda_m}{4(\gamma-1)}\left(\frac{E_\gamma}{M}\right)\left(\frac{r_s}{r_\gamma}\right)^3\left(\frac{\rho_c}{\rho_0}\right).
\label{aeden}
\end{eqnarray}
These represent the characteristic temperature and energy density of the photon-pair sphere. 
They are plotted in the bottom panel of Fig.~\ref{toten} as functions of 
the initial core centre density $\rho_c(t)$, when homologous collapse starts.
Beside, the photon dimensionless energy $\nu/\langle T_\gamma\rangle$ and distribution 
function $F_\nu(\nu/\langle T_\gamma\rangle)$ in phase space are relativistic 
and cosmological redshift invariants 
resulting from the Liouville theorem (see e.g.~\cite{Ehlers1971}). The photon energy $\nu$ 
has a maximum (or peak energy) at $\nu^{\rm max}_\gamma\propto \langle T_\gamma\rangle$.

We close this section with the following emphases. Instead of photon-pair spheres, photon-pair jets are formed for angular momentum conservation in the cases of rotating stellar core collapse, two stellar cores merger, and the presence of strong magnetic fields. Such a photon-pair jet is in the direction along with the system axial symmetric axis. One can study a photon-pair jet by considering the centrifugal or magnetic potentials with gravitational potential $\phi$ in the local virial theorem (\ref{dgravP}). The physical quantities should be functions of 
radius $r$ and azimuth angle $\theta$, characterised by not only the initial core centre density $\rho_c(t)$, but also photon-pair 
jet collimation angle $\theta_{\rm coll}$ determined by system angular momenta and magnetic fields.

\section{Intrinsic correlations and connections with GRB sources}
\label{scaling}

We have just shown the possible mechanism of how the vast gravitational energy converts to the huge energy of an opaqued photon-pair sphere energy in homologous core collapses. In the case that 
photon-pair sphere heat energy is much larger than its baryon loading, such an energetic photon-pair sphere/jet undergoes ultra-relativistically hydrodynamic outward expansion, and its temperature and radius thickness remains approximately $T_\gamma$ and $R_\gamma$ due to relativistic effects. At the transparency point, the photon-pair sphere becomes a photosphere. 
These were studied in the literature, and some details of using hydrodynamic 
equations and rate equation of back and forth process $\gamma+\gamma \leftrightarrow e^++e^-$ 
in general relativity framework can be found in Refs.~\cite{Ruffini1999,Ruffini2000,Ruffini2003}. 
During relativistic expansion, photon-pair sphere/jet proceeds ``internal'' energy dissipation processes, it sweeps and interacts with the circumburst media executing the ``external'' energy dissipation processes, accounting for GRB phenomena.  
Thus, the photon-pair sphere/jet kinematic motions (baryon loading and collimation), 
dynamical interactions (collisions, shocks and fluid dynamics) and 
energy dissipation mechanism (particle accelerations and radiation mechanisms) 
are important and have been modelled to account for the complex phenomena of GRB sources observed, see for example Refs.~\cite{2004RvMP...76.1143P,2006RPPh...69.2259M,2014ARA&A..52...43B,2015JHEAp...7...73D,2015PhR...561....1K,zhang_2018,2019Univ....5..110R,Ruffini2019}.
The opaqued photon-sphere formation is only the initial phase of GRB engines (progenitors). Therefore our present study is not in the position of directly explaining complex phenomena of GRB sources observed. 
Nevertheless, we examine photon-pair sphere properties and possible connections with GRB observations in this section.  

\subsection{Photon-pair sphere characteristics}

The total core mass $M$, initial core centre density $\rho_c(t)$, and 
eigenvalue $(\gamma,\lambda_m)$ uniquely represent different configurations 
of homologously collapsing core,
\begin{eqnarray}
\left\{M,\rho_c(t),(\gamma,\lambda_m)\right\}.
\label{irhoc}
\end{eqnarray}
The configuration determines the homologous core density profile $f^3(r)$, 
outer boundary radius $r_s$ (\ref{rs}) and the photon 
sphere radius $R_\gamma$ (\ref{para1}), see Eq.~(\ref{meand}), 
as well as all other baryon core and photon-pair sphere 
quantities we have obtained. The various configurations (\ref{irhoc}) 
in fact correspond to different photon-pair spheres.

From the results presented in Fig.~\ref{toten}, 
we find the ranges of the total photon-pair sphere energy 
$E_\gamma\sim (10^{-2}\sim 10^{-4}) M$ and number $N_\gamma\sim (1\sim 10^{-3})N$, 
taking into account the centre sound velocity $(v^c_s)^2$ variation, 
see right of Fig.~\ref{vrho}. These results are consistent with the total energy and entropy budgets. They are necessarily required to explain the cosmological origin of GRB phenomena observed. On the other hand, recall that the variation of 
gravitational energy $GM^2/(2R)$ is $M/4$ for a core-collapse from infinity $R=\infty$ 
to $R=2GM$. It implies that the total photon-pair sphere energy $E_\gamma$ is only a few per cent of maximally available gravitational energy $M/4$. The rest 
could be for core hydrodynamical and kinematic motion, 
as well as gravitational wave (non-spherical collapse) 
and other particles emissions. Such conversion from the available gravitational energy to the photon-pair sphere energy occurs in a couple of seconds. It means that the gravo-thermal dynamics is very efficient via
hadron collisional relaxation and photon production in gravitational collapses.

Depending only on the averaged core thermal index $1\lesssim \gamma\lesssim 4/3$, 
Figures \ref{ted}, \ref{ndensize}  and \ref{toten}  represent the qualitative results and 
characteristic properties of the photon-pair sphere, namely the photon temperature 
$T_\gamma$, energy density $\rho_\gamma$ 
and number density $n_\gamma$, as well as the photon-pair sphere 
size $R_\gamma$ and opacity. These results show that the photon-pair sphere is deeply 
opaque, highly energetic and rather sizeable.
Its temperature, size, energy and number densities are in the correct ranges, necessarily 
for 
the initial configurations of the opaque photon-pair sphere,
i.e., {\it fireball or fireshell}, that subsequently undergoes various hydrodynamical 
and electromagnetic evolutions 
to explain GRB phenomena observed. 

Notwithstanding, we show that the gravo-thermal dynamics can necessarily account for 
main energetic natures of GRB progenitor dynamics. 
In general, the realistic situations of gravitational collapse, baryon collisional relaxation and photon production are much more complicated. It is expected that the photon-pair sphere energy and number densities are reduced, 
the energy spectrum is softened, and the collapsing time is prolonged. 
Among others, the main reasons are 
that internal perturbation and/or shock waves are formed due to the baryon matter EoS 
(\ref{eos}) variations, phase transition and hydrodynamics, as well as the presence of 
macroscopic electromagnetic fields and thermal photon-pair sphere dynamics, even general 
relativistic effects. These are beyond the scopes of this article and subjects for 
future studies.

\subsection{Universal scaling laws with only one index parameter}\label{uscaling}

All physical quantities of photon-pair sphere 
are described in the rest frame anchored to the origin of homologously collapsing core. The photon-pair sphere size $R_\gamma$ (\ref{Rsize}), total energy $E_\gamma$ (\ref{tote}), mean temperature $\langle T_\gamma \rangle$ (\ref{atem}) and energy density $\langle \rho_\gamma \rangle$ (\ref{aeden})
depend on the core centre density $\rho_c(t)$ (\ref{irhoc}) when homologous 
core collapse starts. 
Therefore they are intrinsically related each others via 
$\rho_c(t)$. Equation (\ref{atem}) gives $\langle T_\gamma \rangle\propto (\rho_c)^{\gamma-1}$, and
Eq.~(\ref{tote}) gives $E_\gamma\propto (\rho_\gamma/\rho_c)\propto \langle T_\gamma \rangle^4/\rho_c
\propto \rho_c^{(4\gamma -5)}$. Recall that the initial core centre density $\rho_c(t)$ values (\ref{irhoc}) represent different GRB sources. As a result, eliminating the dependence on $\rho_c(t)$, we qualitatively obtain the intrinsic 
correlation or universal scaling law between the photon-pair sphere total energy $E_\gamma$ and characteristic 
energy scale (temperture) $\langle T_\gamma \rangle$
\begin{eqnarray}
E_\gamma \propto \langle T_\gamma \rangle^{\chi}, \quad \chi= \frac{4\gamma -5}{\gamma-1}>0. 
\label{chi}
\end{eqnarray}
This correlation (\ref{chi}) and those below are universal for 
all photon-pair spheres in the scenes that they are independent of the homologously collapsing core mass $M$ and initial core centre density $\rho_c$(t) in Eq.~(\ref{irhoc}).
Here the theoretical index $\chi$ depends on the averaged value of effective thermal index $\gamma$. 
The present theoretical model and EoS (\ref{eos}) are too simple and preliminary to quantitatively determine the averaged value of effective thermal index $\gamma$. Nevertheless, we find 
that $\gamma\sim {\mathcal O}(1)$ values are on the right spot to give $\chi\sim {\mathcal O}(1)$ in Eq.~(\ref{chi}). We will treat 
$\chi\sim {\mathcal O}(1)$ as a unique parameter for all correlations discussed here.

From the viewpoint of total photon-pair sphere energy conservation, 
we can define the total photon-pair sphere luminosity  
\begin{eqnarray}
L_\gamma \equiv 4\pi  c R_\gamma^2\langle \rho_\gamma\rangle \propto \langle \rho_\gamma\rangle  \rho_c^{-(2-\gamma)}\propto \rho_c^{5\gamma-6}\propto \langle T_\gamma \rangle^{(\chi+1)}.  
\label{chil}
\end{eqnarray} 
This givs rise to the correlation of photon-pair sphere luminosity and temperature. 
The correlating index is not a new parameter, but $(\chi+1)$ relating to the 
index $\chi$ in the correlation (\ref{chi}). The $E_\gamma-T_\gamma$ (\ref{chi}) and $L_\gamma-T_\gamma$ (\ref{chil})
correlations lead to $E_\gamma-L_\gamma$ correlation, 
\begin{eqnarray}
E_\gamma \propto L_\gamma^{\frac{\chi}{\chi+1}}.  
\label{chile}
\end{eqnarray}
With only one free parameter $\chi\sim {\mathcal O}(1)$, three correlations $E_\gamma-T_\gamma$ (\ref{chi}), 
$L_\gamma-T_\gamma$ (\ref{chil}) and $E_\gamma-L_\gamma$ (\ref{chile}) 
coherently reflect the energetic natures of gravo-thermal dynamics, which possibly
explains GRB energetics. 

Moreover, the photon-pair sphere size $R_\gamma$ (\ref{Rsize}) represents the characteristic time scale $\tau_\gamma=R_\gamma/c$ of the photon-pair sphere. It should relates the temporal duration of GRBs. Equation (\ref{Rsize}) gives $R_\gamma\propto  \rho_c^{-(2-\gamma)/2}$ dependence on the core centre density $\rho_c(t)$. 
Using the $E_\gamma$, $T_\gamma$ and $L_\gamma$ relations
to the core center density $\rho_c(t)$, 
we obtain three anti-correlations $T_\gamma-\tau_\gamma$, $E_\gamma-\tau_\gamma$ and $L_\gamma-\tau_\gamma$:
\begin{eqnarray}
\langle T_\gamma \rangle &\propto& \tau_\gamma^{-\delta}=\tau_\gamma^{-\frac{2}{3-\chi}},\label{time1}\\
E_\gamma &\propto& \tau_\gamma^{-\chi\delta}=\tau_\gamma^{-\frac{2\chi}{3-\chi}},\label{time2}\\
L_\gamma &\propto& \tau_\gamma^{-\delta(\chi+1)}=\tau_\gamma^{-\frac{2(\chi+1)}{3-\chi}},
\label{time3}
\end{eqnarray}
where
\begin{eqnarray}
\delta\equiv 
2\frac{\gamma -1}{2-\gamma}= \frac{2}{3-\chi}>0.
\label{delta}
\end{eqnarray}
These are consistent with correlations (\ref{chi}), 
(\ref{chil}) and (\ref{chile}), and indeed
$E_\gamma\propto L_\gamma \tau_\gamma$ as it should be.  
Three anti-correlations $T_\gamma-\tau_\gamma$, $E_\gamma-\tau_\gamma$ and $L_\gamma-\tau_\gamma$ coherently reflect the time-scale natures of gravo-thermal dynamics 
for GRB energetics. 

Via the sound velocity $(v^c_s)^2$ (\ref{soundc}) at the core center, the maximal photon-pair sphere temperature $T_\gamma^{\rm max}\propto (\rho_c)^{\gamma-1}$ (\ref{tmax}). Its dependence on the core center density $\rho_c$ is in the same way as the mean temperature 
$\langle T_\gamma \rangle\propto (\rho_c)^{\gamma-1}$ (\ref{atem}), and  $T_\gamma^{\rm max}\approx 6.87 \langle T_\gamma \rangle$. 
Therefore the correlations between $T_\gamma^{\rm max}$ and other physical quantities can be obtained by substituting 
$T_\gamma^{\rm max}$ for $\langle T_\gamma \rangle$ 
in Eqs.~(\ref{chi}), (\ref{chil}) and (\ref{time1}). 
These could be of some interests. Because the maximal photon-pair sphere temperature $T_\gamma^{\rm max}$ might be related to the high-energy threshold of GRB photons. 

The theoretical correlations (\ref{chi}), (\ref{chil}) and (\ref{chile}), 
anti-correlations (\ref{time1}), (\ref{time2}) and (\ref{time3}) 
are not completely independent each other. However they base on only 
one free parameter $\chi\sim {\mathcal O}(1)$. 
They should receive corrections from the cosmological redshift $z$, 
$E_\gamma \rightarrow E_\gamma/(1+z)$, $\langle T_\gamma \rangle \rightarrow \langle T_\gamma \rangle/(1+z)$, $\tau_\gamma \rightarrow \tau_\gamma (1+z)$ and 
$L_\gamma \rightarrow L_\gamma$. 

\subsection{Possible connections to GRB observations}

The theoretical correlation (\ref{chi}) seems to be the empirical Amati/Yonetoku relation $E_{\rm iso} \propto E_p^2$ in GRBs \cite{Amati2002,Yonetoku2004}, if we identify the photon-pair sphere energy $E_\gamma$ and mean energy $\langle T_\gamma \rangle$ to the observed total isotropic energy $E_{\rm iso}$ and peak energy $E_p$, or $\nu_\gamma^{\rm max}$ at which the maximum of photon energy spectrum $\nu F_\nu$ locates. However, we cannot simply 
make such identification for the reasons that the 
photon-pair sphere undergoes complex processes up to the transparency at which it becomes an observably relevant photosphere. One of reasons is that the peak energy $E_p$ value sensitively depends on the photon-pair sphere baryon loading, see for example, Refs.~\cite{Ruffini1999,Ruffini2000,Meszaros2000,Zhang2021}.  
We attempt to give a general discussion on the possible connections between theoretical correlations (\ref{chi}-\ref{delta}) of different photon-pair spheres and observational correlations of different GRB sources.

The photon-pair spheres provide the necessary energy budgets for GRB events. Suppose that each photon-pair sphere created by a gravitational collapse process corresponds to a GRB event/source observed. As aforementioned at the beginning of the section, a photon-pair sphere ({\it initial state}) undergoes complex ({\it intermediate processes}) to a ({\it final state}) of GRB phenomena observed. The photon-pair sphere quantities $T_\gamma$, $E_\gamma$, $L_\gamma$ and $R_\gamma$ are not exactly the same as observed GRB event's mean energy, total energy, total luminosity, and time duration. Therefore, the correlations (\ref{chi}-\ref{delta}) cannot be directly examined by observed GRB events. However, photon-pair spheres' overall quantities ($T_\gamma,E_\gamma, L_\gamma,
R_\gamma$) and basic relations (\ref{chi}-\ref{delta}) must imprint themselves in GRB data measured. Though the {intermediate processes} from a photon-pair sphere to an observed GRB source are complex, they are randomly different from one GRB source to other. Such randomness implies that the data of GRB events relevant to overall photon-pair sphere quantities should scatter around the basic relations (\ref{chi}-\ref{delta}) if these scaling relations truly reflect the universal natures of centre engines for GRB energetics. The more GRB events are taken into account, the more statistically confident level will be achieved on the validity of the basic relations (\ref{chi}-\ref{delta}) of photon-pair spheres. 
However, these are very general discussions. The relevant GRB phase identification, spectrum and event selection, elaborate data analysis are required to give conclusions. These are subject to future studies. If these theoretical correlations and scaling laws are correct and verified, 
GRB sources can be considered as standard candles for determining cosmological distance.

\comment{
It is worthwhile to examine these theoretical correlations and index relations from observational data. There have been 
many empirical studies on Gamma-ray burst prompt correlations, 
see some more details and references in Refs.~\cite{Dainotti2016}.}

\section{Thermodynamic of photon-pair sphere formation}\label{form}

It is necessary to study the thermodynamic of photon-pair sphere formation in gravitation collapses. It gives further insight into the dynamics of baryon relaxation 
(virial theorem) and photon-pair sphere formation in the gravo-thermal 
dynamics during gravitational collapses.
It is very different from the hydrodynamical evolution of the photon-pair sphere.  

\subsection{Negative pressure and gravitational energy gain}

The first thermodynamics law for the adiabatic transformation
of the system, 
in which the particle number changes in time, is given by 
\cite{Prigogine1988,Prigogine1989},   
\begin{eqnarray}
dQ=d(\varrho V) +{\mathcal P} dV -\frac{\varrho + {\mathcal P}}{\tilde n} d(\tilde n V), \quad dQ=0.  
\label{tdnt}
\end{eqnarray}
The system is of the volume $V$, the particle number density $\tilde n$, usual internal energy density $\varrho$ and pressure 
${\mathcal P}$. In Eq.~(\ref{tdnt}), the third term of the negative sign represents the system gains energy. It is due to the change in the particle number $\tilde nV$. 
The thermal pressure ${\mathcal P}$ is determined by the energy production $d\varrho$ and particle production $d\tilde n$. Equation (\ref{tdnt}) is equivalent to $d\varrho=(\varrho+{\mathcal P})d\tilde n/\tilde n$ or ${\mathcal P}=(\tilde nd\varrho -\varrho d\tilde n)/d\tilde n$.

As the time-scale hierarchy (\ref{Thy}) discussed in Sec.~\ref{Homo}, the macroscopic 
gravitational collapse and hydrodynamic processes are approximately adiabatic, w.r.t. the 
microscopic processes for creation and thermalization of particle energy and the number 
distributions. We apply the thermodynamics law (\ref{tdnt}) to the system that undergoes gravitational collapse and photon-pair sphere formation. The self-gravitating system contains 
(i) the baryon core of density $n$ and conserved baryon number $N$; 
(ii) the photon-pair sphere of density $n_\gamma$ and increasing photon number $N_\gamma$. 
The photon production is caused by hadron collisions, consuming the hadron 
``heat'' energy $dF$ (\ref{dgravP}). The total internal energy density 
$\varrho=F/V+\rho_\gamma$ and pressure ${\mathcal P} =p+p_\gamma$. 
The photon-pair sphere is opaque, and no heat energy is transferred outside 
the system $dQ=0$. Considering the first thermodynamics law in the presence of 
gravitational potential energy, see for example \cite{Koenig1936}, 
we generalise the law (\ref{tdnt}) to
\begin{eqnarray}
dQ=d(\varrho V) +{\mathcal P} dV -\frac{\rho_\gamma + p_\gamma}{n_\gamma} d(n_\gamma V) + dU, \quad dQ=0.  
\label{tdntg1}
\end{eqnarray}
This is the total energy conservation of the system. The baryon core and photon-pair sphere exchange heat energy. This can be seen by rewriting Eq.~(\ref{tdntg1}) as  
\begin{eqnarray}
-d(\rho V) - p dV - dU = d(\rho_\gamma V) + p_\gamma dV -\frac{\rho_\gamma + p_\gamma}{n_\gamma} d(n_\gamma V).  
\label{tdntg2}
\end{eqnarray}
The left-handed side indicates the hadron collision energy pumped into the photon-pair sphere. The right-handed side can be effectively 
rewritten as $d(\rho_\gamma V) +(p_\gamma+p_n) dV$ with a negative pressure $p_n$ defined as
\begin{eqnarray}
p_n\equiv -\frac{\rho_\gamma +p_\gamma}{n_\gamma} \frac{d(n_\gamma V)}{dV} < -(\rho_\gamma +p_\gamma) = -4 p_\gamma ,  
\label{negp}
\end{eqnarray}
where photon production leads to $d(n_\gamma V)>0$ and $dn_\gamma >0$. 

In contrast with the normal positive photon pressure $p_\gamma$, 
negative pressure $p_n$ is in energetic favor of gravitational collapse. 
The total effective photon pressure $(p_\gamma+p_n)< - 3p_\gamma$ is negative. 
Therefore, the photon-pair sphere creation of increasing photon number density 
$dn_\gamma >0$ is energetically favourable in gravitational collapse. 
It gains energy from the hadron collisional energy, which comes from the gravitational potential energy. 
This result indicates that the total effective heat function of the photon-pair sphere 
$h_\gamma^{\rm eff}\equiv h_\gamma +p_n=\rho_\gamma +p_\gamma +p_n$ can be negative, 
favouring gravitational collapse rather than positively repelling as usually expected. However, when the total photon number stops increasing $d(n_\gamma V)=0$, 
the photon-pair sphere becomes a normal relativistic fluid of positive pressure 
$p_\gamma=\rho_\gamma/3$, pushing outwardly against gravitational attraction.   
In this article, we do not consider these photon-pair sphere effects on homologous collapse dynamics since the photon-pair sphere energy density 
$\rho_\gamma$ is much smaller than the baryon core energy density.  

\subsection{Entropy increases in particle relaxation 
and production}

We turn now to the second law of thermodynamics. 
To evaluate the entropy flow and the entropy production, 
one starts from the total differential of the entropy ${\mathcal S}$
\cite{Prigogine1988,Prigogine1989},   
\begin{eqnarray}
{\mathcal T}d{\mathcal S}=d(\varrho V) +{\mathcal P} dV -\mu d(\tilde n V), 
\label{entropy0}
\end{eqnarray}
where ${\mathcal T}$ and 
$\tilde n \mu = \varrho + {\mathcal P}- {\mathcal T}{\mathcal S}/V$ 
are the temperature and chemical potential respectively. 
Applying the total differential of entropy (\ref{entropy0}) to the 
self-gravitating system of the baryon core and photon-pair sphere, 
we have,  
\begin{eqnarray}
TdS +T_\gamma dS_\gamma= d(\rho V) + p dV
+d(\rho_\gamma V) + p_\gamma dV -\mu_\gamma d(n_\gamma V), 
\label{tdntg3}
\end{eqnarray}
where $T$ and $S$ are baryon core temperature and entropy, 
$T_\gamma$ and $S_\gamma$ are photon-pair sphere temperature and entropy. 
The photon-pair sphere chemical potential $\mu_\gamma$ is
\begin{eqnarray}
\mu_\gamma = n_\gamma^{-1}\Big(\rho_\gamma + p_\gamma- T_\gamma S_\gamma/V\Big). 
\label{phchemical}
\end{eqnarray}
Substituting the first law (\ref{tdntg2}) to the total differential of entropy (\ref{tdntg3}), we obtain the total entropy increase
\begin{eqnarray}
TdS +T_\gamma dS_\gamma= -dU +\frac{T_\gamma S_\gamma}{n_\gamma V}d(n_\gamma V) >0, 
\label{Sinc}
\end{eqnarray}
since $-dU >0$ and $d(n_\gamma V) >0$. Assuming the baryon entropy 
$TdS= -dU$ purely comes from gravitation energy conversion, 
we have $dS_\gamma= S_\gamma d(n_\gamma V)/{n_\gamma V}$. 
This leads $S_\gamma\propto n_\gamma V$, namely photon entropy 
produced is proportional to photon number produced.

We thus conclude that in gravitational collapses, the processes of baryon relaxation and photon-pair sphere formation are energetically and entropically favourable

\section{Conclusion and remarks}\label{end}

Using a simplified model describing homologous gravitational collapses of massive and dense stellar baryon cores, we present a 
preliminary understanding of how and why
gravo-thermal catastrophe occurs in gravitational collapses, 
converting gravitational potential energy to observable photon energy. 
Such gravo-thermal dynamics attributes to two aspects.
(i) From self-gravitating core potential energy, baryons gain their heat energy via collisional relaxation. We adopt the virial theorem to approximately describes this process. 
(ii) Via baryon collisions, photons are produced and gain their energy from baryons' heat energy. We approximately calculate this process by using the photon production by heavy-ion collisions. The vast difference between macroscopic 
and microscopic time scales, i.e., the time scale hierarchy (\ref{Thy}), 
is {\it a priori} the prerequisite for using semi-analytical analyses at the qualitative level. 
The obtained physical results need verifications for theoretical 
self-consistencies {\it a posteriori} and compared with observational data. 
As a result, we show that 
a photon-pair sphere  formation process is 
energetically and entropically favourable. 
Such created photon-pair spheres possess fundamental physical quantities, i.e., size, 
total energy, energy and number densities that possibly qualitatively account for 
the main energetic features of GRB progenitors. 
All these quantities uniquely depend on the baryon core centre density 
$\rho_c(t)$ (\ref{irhoc}), which is determined by homologous core mass $M$ 
and radius $r_s$, as well as the eigenvalue $(\gamma,\lambda_m)$ 
of homologous configurations. These configurations represent different GRB 
sources (events). There are two basic parameters in the baryon core EoS (\ref{eos}): 
the averaged thermal index $1<\gamma<4/3$ and maximal core central density 
$\rho^{\rm max}_c(\gamma,\kappa)\gtrsim \rho_0$ (\ref{amini}). 
Eliminating the $\rho_c(t)$-dependence in these physical quantities, we obtain 
intrinsic correlations (universal scaling laws) among them, 
in terms of only one parameter $\chi=\chi(\gamma)\sim {\mathcal O}(1)$. 
These theoretical correlations (scaling laws) and their scaling indices 
must confront observational data.

In reality, the gravo-thermal dynamics for stellar core collapse and photon-pair sphere formation must be very complex. There are not only violent hadron collisions 
and photon productions, but also violently strong electromagnetic field fluctuations, 
plasma oscillations, and hadron-quark phase transition in microscopic scales. 
As a consequence, the thermal index $\gamma$ in the effective EoS (\ref{eos}) 
should be a function of space and time. Such $\gamma$-inhomogeneity 
should cause sound velocity (\ref{sound}) variation and shock wave occurrence, 
impacting on homologous collapse and photon-pair sphere formation. Indeed, the macroscopic 
hydrodynamics of baryon core and photon-pair sphere, on the other hand, 
play important roles in the gravo-thermal dynamics. Due to 
the vast time scale hierarchy (\ref{Thy}), it is very inviting to adopt 
analytical approaches to microscopic processes and 
numerical simulation algorithms for macroscopic processes, to give a quantitative
study of the gravo-thermal dynamics in connection with GRB observations.

We end this article by making some remarks on possible relevant issues of neutrino bursts, high-energy photons, cosmic rays and high-frequency gravitation wave emissions. In the photon-pair sphere, the electron-positron pair annihilation $e^++e^-\rightarrow \nu+\bar\nu$ is the dominant neutrino production, and the neutrino opacity is mainly due to
the scattering $\nu+e^\pm\rightarrow \nu+e^\pm$. If neutrinos are trapped inside the photon-pair sphere, the neutrino mean  energy $\langle E_\nu\rangle$
is about $T_{e^+e^-}\approx T_\gamma$. At decoupling bursts, neutrino sphere total energy $E^\nu_{\rm tot}=3.1\times 10^{51}{\rm erg} [E_{\gamma 52}R_{\gamma 6.5}]^{11/16}$ and
mean energy $\langle E_\nu\rangle =56\,{\rm MeV}\times [E_{\gamma 52}]^{1/4}[R_{\gamma 6.5}]^{-3/4}$ \cite{Koers2005}, where $E_{\gamma 52}=E_{\gamma}/10^{52} {\rm erg}$ and $R_{\gamma 6.5}=R_{\gamma}/10^{6.5}{\rm cm}$. They are relevant for observations and should follow the same scaling laws discussed in Sec.~\ref{uscaling}. If neutrinos are not trapped, neutrino continuous emission and cooling processes occur, which may prevent the photon-pair sphere from forming. This does not seem to be the case. The reasons are that (i) electron-positron ($e^+e^-$) plasma density and size are large enough, see Fig.~\ref{ndensize}, to trap neutrinos; (ii) the hadronic photon production rate (\ref{gammaNR}) is fast enough, namely, 
its time scale (\ref{prodt}) is shorter than the time scale (\ref{Tgamma}) of $e^+e^-$ pair production. 
Moreover, due to its weak interaction nature, the $\bar\nu \nu$ production time scale is longer than $e^+e^-$ pair production. 
However, these reasons require a quantitative analysis based on the time scale hierarchy (\ref{Thy}) 
and the time scale of 
photon-pair sphere hydrodynamical expansion. It is worthwhile to mention that in the photon-pair sphere, hot and dense 
photon-electron-positron plasma modes emit high-frequency 
($\omega\sim T_\gamma$) gravitational wave 
\cite{1972gcpa.book.....W, Han2014}.

\comment{We estimate the gravitational wave frequency $\omega\sim T_\gamma$,  total flux $dE_{\rm grav}/dt \sim (GT_\gamma^2) E_\gamma^2\approx 10^{-38}(T_\gamma/m)^2E_\gamma^{52} $ and total energy $E_{\rm grav}\sim (GT_\gamma^2) E_\gamma^2R_\gamma \approx 1.59\times 10^{36} 10^{52} {\rm erg} (T_\gamma/m)^2 E_{\gamma 52} (E_{\gamma 52}R_{\gamma 6.5})(10^{52} erg 10^{6.5} cm)$. 
}

Last, but certainly not least, we discuss the possibilities 
for the continuous emissions of ultra-high-energy cosmic rays (protons), ultra-high-energy photons 
and neutrinos from the photon-pair sphere.  
Due to the gravitational attraction and hydrodynamical expulsion, the photon-pair sphere is split into two parts. The inner part of the energy 
$E_\gamma^{\rm in}\lesssim E_\gamma/2$ falls inward, while its outer part expands outward, and the separatrix radius locates at $R_{\rm sep}\approx 4GM$ \cite{Ruffini2003}. As mentioned, the expanding outer part leads to complex GRB phenomena. While the falling inner part takes a long time to trap onto the horizon $R_+=2GM$ because of the general relativity effect. 
The inner part energy density $\rho_\gamma
\sim T_\gamma^4> m_e^4$ indicates violent fluctuating electromagnetic fields $\tilde E_{\rm em}>E_c=B_c=m_e^2/e$. These fields possibly accelerate protons to ultra high energies $\mathcal E\approx e \tilde E_{\rm em} \ell$, up to 
GeV, TeV, PeV, EeV and $10^{21}$ eV, in small lengths $\ell < (\mathcal E/m_e)\lambda_e$ and times $\tau < (\mathcal E/m_e)\tau_e$, where the Compton length $\lambda_e\sim 10^{-11}$cm and time $\tau_e\sim 10^{-21}$sec. These high-energy protons can be candidates of high-energy cosmic rays. They 
produce high-energy photons via possible QED processes by interacting with charge particles and external classical magnetic fields. They produce high-energy neutrinos via proton-proton collisions. 
The inner part size and optical depth are much smaller than the outer part. The inner part activities of releasing particles and energy are continuous, in contrast to the outer part decoupling bursts. 
It may account for the high-energy cutoff power-law spectrum of GRBs. The inner part activities can be observed after the outer part becomes transparent. All these speculations need further studies.

 \comment{
the virial process of "violent relaxation" and virial theorem, 
a good explanation and estimation
, see page 328, bottom, and reference there, The book of The Universe by KOLB  
GRBs' energy $E_{\gamma,{\rm iso}}\sim 10^{49-54}$ergs }

\section
{\bf Acknowledgment}  
The author thanks Professor Ruffini and ICRANet members for many discussions 
on GRB physics and phenomena.

\section{Appendix: the time evolutions of homologous profiles}\label{app}

\begin{figure}[ht]
\includegraphics[width=3.0in]{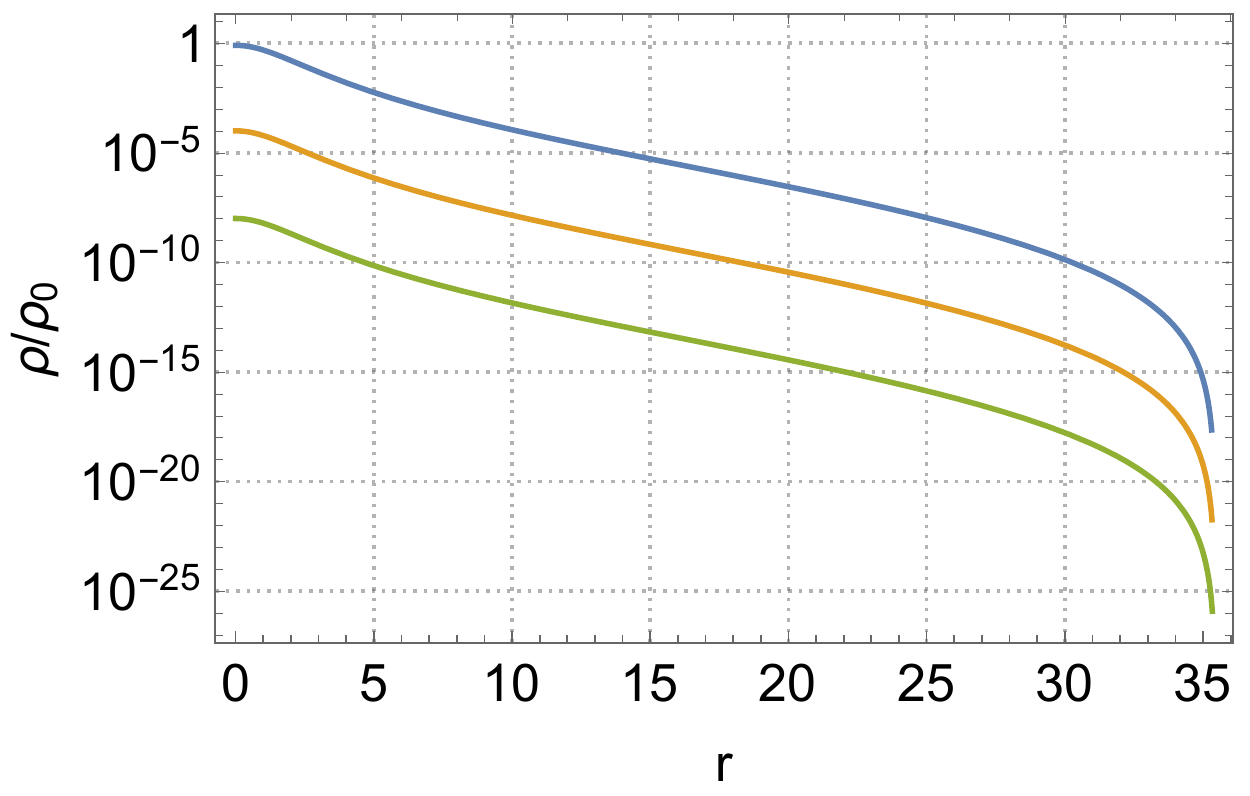}
\includegraphics[width=3.0in]{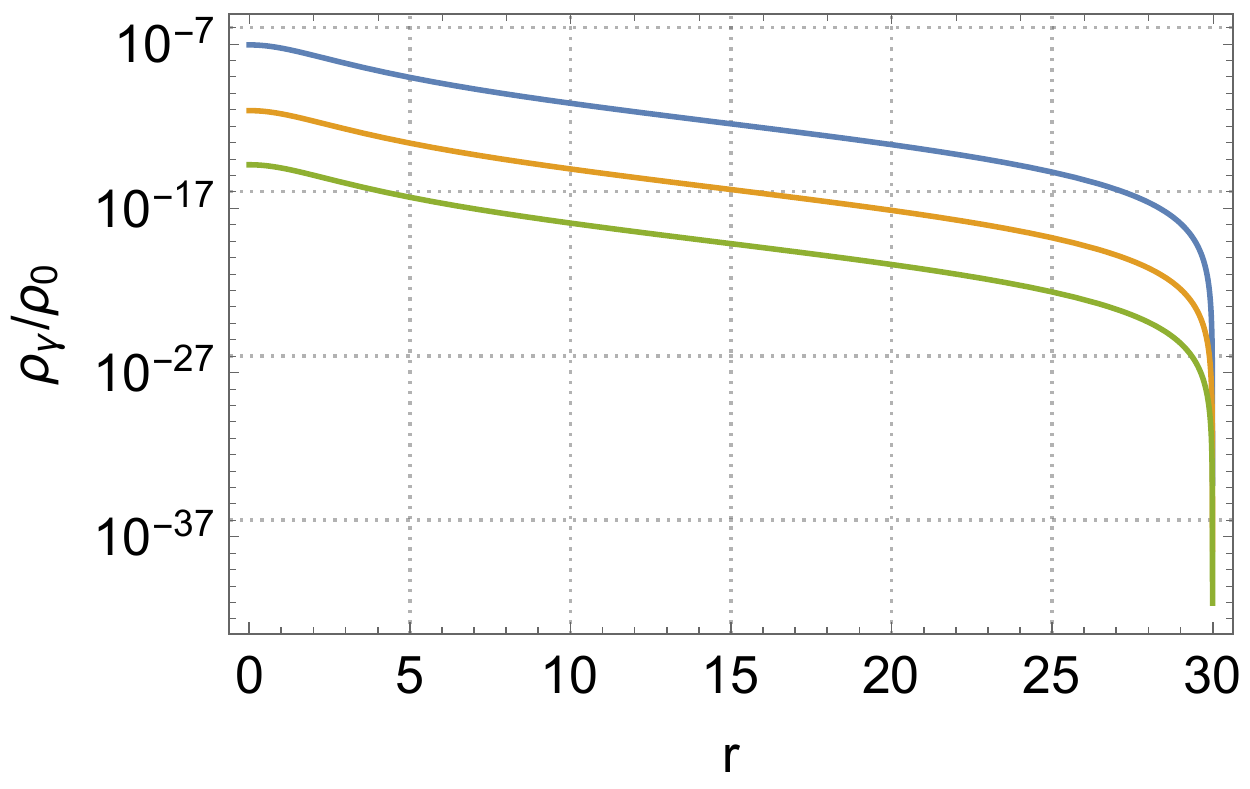}
\includegraphics[width=3.0in]{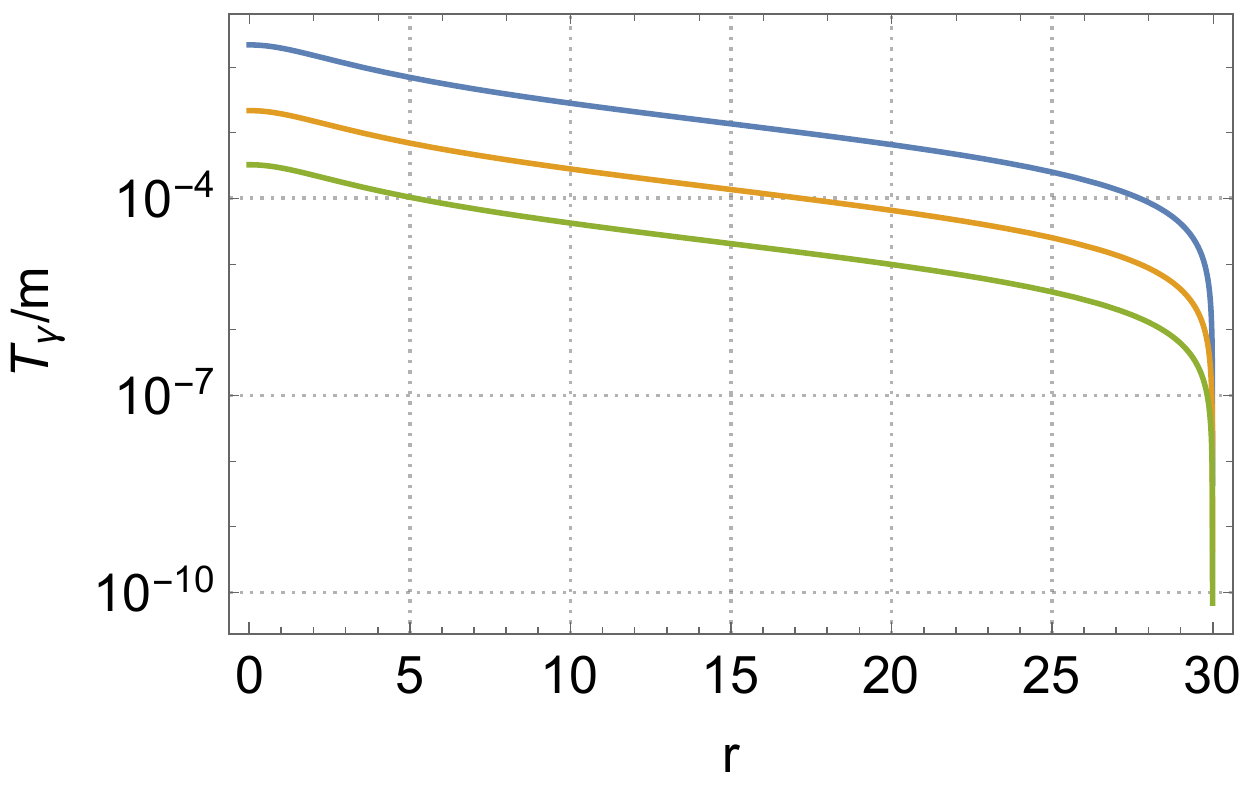}
~~~~~\includegraphics[width=3.0in]{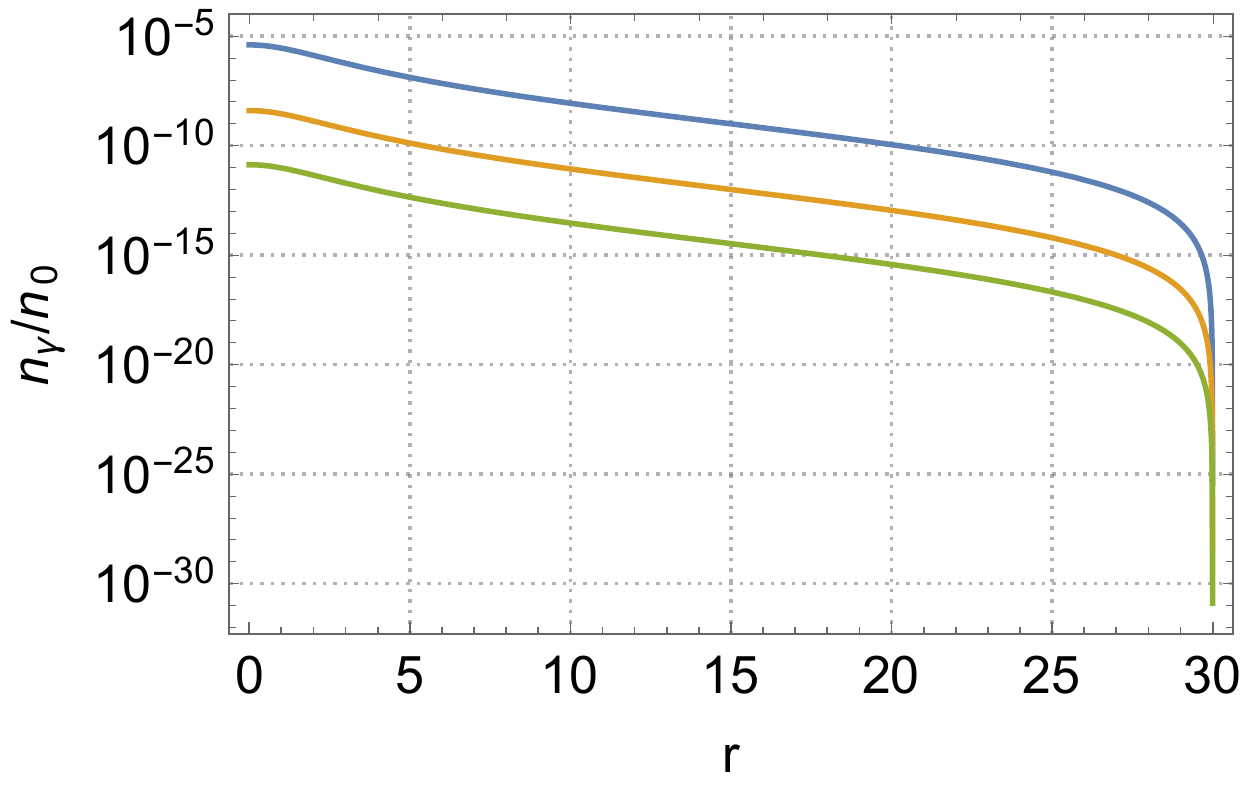}
\caption{Colours on line. The green (below) line for time $t_1/\tau_{\rm min}\approx 10^6$, orange (middle) line for time $t_2/\tau_{\rm min}\approx 10^4$ and blue (above) line for time $t_3/\tau_{\rm min}\approx 10^2$, see Fig.~\ref{vrho}. As described in text, the time sequence $t_1>t_2>t_3$ for ongoing gravitational collapses is in the 
inverse direction of the time arrow, from $t\gg \tau_{\rm min}$ to $t=\tau_{\rm min}$.
These figures show the time evolutions of homologous profiles of 
baryon core density, photon-pair sphere temperature, energy and number densities for the eigenvalue $(\gamma,\lambda_m)=(1.23,8\times 10^{-5})$.  Top panel: the baryon core density $\rho(r,t)/\rho_0$ (\ref{den1}) (left), and photon energy density 
$\rho_\gamma(r,t)/\rho_0$ (\ref{gammaEd}) (right). Bottom panel: the photon temperature $\langle T_\gamma(r,t)\rangle/m$ (\ref{dgravT1}) (left) and number density $\langle n_\gamma\rangle /n_0$ (\ref{gammaNd}) (right). 
Note that typic baryon mass $m\approx 1$ GeV, the nuclear saturation density 
	$\rho_0\approx 2.4\times 10^{35} {\rm ergs}/{\rm cm}^3$ and  
 $n_0\approx 1.6\times 10^{38}/{\rm cm}^3$. 
}\label{timev}
\end{figure}

For readers' convenience, we show in Fig.~\ref{timev} the time evolutions of homologous profiles of baryon core density, photon-pair sphere temperature, energy and number densities. These homologous profiles do not change their shapes in time. However, their amplitudes change greatly in time as the gravitational collapse process proceeds. Note that the homologous core radius $r_s$ and 
photon-pair sphere radius $r_\gamma$ are independent of the collapsing time $t$. They depend on 
the eigenvalue $(\gamma,\lambda_m)$, core centre mass $\rho_c$ 
and homologous core mass $M$, see Eq.~(\ref{meand}). Moreover, Figure \ref{timev} gives us some intuitive ideas about how these homologous profiles smoothly evolving and varying in macroscopic time and length scale. Whereas the microscopic processes of 
baryon collisional relaxation, photon production and thermalisation violently 
take place in tiny spacetime shells of width $\tau_{\rm micro}$ 
and $\ell_{\rm micro}\approx c \tau_{\rm micro}$.

\providecommand{\href}[2]{#2}\begingroup\raggedright\endgroup

\end{document}